\documentclass[12pt,oneside,reqno]{amsart}
\usepackage{amsmath,amssymb,amsthm,textcomp}
\usepackage{amsfonts,graphicx}
\usepackage[mathscr]{eucal}
\usepackage{color}
\usepackage{enumitem}
\usepackage[table]{xcolor}
\usepackage{booktabs}

\usepackage{subfig}
\usepackage{epsfig}
\usepackage{caption}
\theoremstyle{definition}
\usepackage{float}
\usepackage{mathtools} 
\usepackage[colorlinks=true,urlcolor=blue,citecolor=blue,linkcolor=blue,bookmarks=true]{hyperref} 
\usepackage[sort & compress,comma,authoryear]{natbib}
\floatstyle{plaintop}
\numberwithin{equation}{section}

\newcommand{\ncom}{\newcommand}

\ncom{\beq}{\begin{equation}}
	\ncom{\eeq}{\end{equation}}
\ncom{\bea}{\begin{eqnarray*}}
	\ncom{\eea}{\end{eqnarray*}}
\ncom{\beqa}{\begin{eqnarray}}
	\ncom{\eeqa}{\end{eqnarray}}
\ncom{\nno}{\nonumber}
\ncom{\non}{\nonumber}
\ncom{\ds}{\displaystyle}
\ncom{\half}{\frac{1}{2}}
\ncom{\mbx}{\makebox{.25cm}}
\ncom{\hs}{\mbox{\hspace{.25cm}}}
\ncom{\rar}{\rightarrow}
\ncom{\Rar}{\Rightarrow}
\ncom{\noin}{\noindent}
\ncom{\bc}{\begin{center}}
	\ncom{\ec}{\end{center}}
\ncom{\sz}{\scriptsize}
\ncom{\rf}{\ref}
\ncom{\s}{\sqrt{2}}
\ncom{\sgm}{\sigma}
\ncom{\Sgm}{\Sigma}
\ncom{\psgm}{\sigma^{\prime}}
\ncom{\dt}{\delta}
\ncom{\Dt}{\Delta}
\ncom{\lmd}{\lambda}
\ncom{\Lmd}{\Lambda}
\ncom{\Th}{\Theta}
\ncom{\e}{\eta}
\ncom{\eps}{\epsilon}
\ncom{\pcc}{\stackrel{P}{>}}
\ncom{\lp}{\stackrel{L_{p}}{>}}
\ncom{\dist}{{\rm\,dist}}
\ncom{\sspan}{{\rm\,span}}
\ncom{\re}{{\rm Re\,}}
\ncom{\im}{{\rm Im\,}}
\ncom{\sgn}{{\rm sgn\,}}
\ncom{\ba}{\begin{array}}
	\ncom{\ea}{\end{array}}
\ncom{\hone}{\mbox{\hspace{1em}}}
\ncom{\htwo}{\mbox{\hspace{2em}}}
\ncom{\hthree}{\mbox{\hspace{3em}}}
\ncom{\hfour}{\mbox{\hspace{4em}}}
\ncom{\vone}{\vskip 2ex}
\ncom{\vtwo}{\vskip 4ex}
\ncom{\vonee}{\vskip 1.5ex}
\ncom{\vthree}{\vskip 6ex}
\ncom{\vfour}{\vspace*{8ex}}
\ncom{\norm}{\|\;\;\|}
\ncom{\integ}[4]{\int_{#1}^{#2}\,{#3}\,d{#4}}
\ncom{\vspan}[1]{{{\rm\,span}\{ #1 \}}}
\ncom{\dm}[1]{ {\displaystyle{#1} } }
\ncom{\ri}[1]{{#1} \index{#1}}

\newtheorem{theorem}{\bf Theorem}[section]

\newtheorem{proposition}{Proposition}[section]

\newtheorem{example}{Example}[section]

\newtheoremstyle
{remarkstyle}
{}
{11pt}
{}
{}
{\bfseries}
{:}
{     }
{\thmname{#1} \thmnumber{#2} }

\theoremstyle{remarkstyle}

\def\eps{\varepsilon}

\begin{document}
	
	\title{ \large O\lowercase{n} \large G\lowercase{eneralized} \large T\lowercase{ransmuted} \large L\lowercase{ifetime} \large D\lowercase{istribution}}

		\author[Alok Kumar Pandey]{Alok Kumar Pandey$^{1}$}
		\author{Alam Ali$^{1}$}
		\author{Ashok Kumar Pathak$^{1*}$
				\\
				$^{1}$D\lowercase{epartment of} M\lowercase{athematics and} S\lowercase{tatistics}, C\lowercase{entral} U\lowercase{niversity of} P\lowercase{unjab},\\
				B\lowercase{athinda}, P\lowercase{unjab}-151401, I\lowercase{ndia}}.
		\thanks{*Corresponding Author E-mail Address: ashokiitb09@gmail.com (Ashok Kumar Pathak)}

	\vspace*{1in}
	\begin{abstract}
		This article presents a new class of generalized transmuted lifetime distributions which includes a large number of lifetime distributions as sub-family. Several important mathematical quantities such as density function, distribution function, quantile function, moments, moment generating function, stress-strength reliability function, order statistics, R{\'e}nyi and q-entropy,  residual and reversed residual life function, and cumulative information generating function are  obtained. The methods of maximum likelihood, ordinary least square, weighted least square, Cram{\'e}r-von Mises, Anderson Darling, and Right-tail Anderson Darling are considered to estimate the model parameters in a general way. Further, a well-organized  Monte Carlo simulation experiments have been performed to observe the behavior of the estimators.
		Finally, two real data have also been analyzed to demonstrate the effectiveness of the proposed distribution in real-life modeling.
		
	\end{abstract}
	
	\maketitle
	
	\section{Introduction} 
	
	In statistics, a number of continuous distributions have been introduced in the literature for model lifetime data in various fields, including medical, insurance, engineering, finance, agriculture, environmental, and biological. Exponential and Weibull distribution are two popular lifetime distributions that have received great attention in the past. These distributions accommodates different shapes of hazard function including increasing, decreasing and bathtub and are extensively used for modeling lifetime data. The generalization of these distributions through parameter addition, random variable transformation, power transformation, and function composition techniques have also been proposed which provide more flexibility in modeling the lifetime data as compared to these distributions.

	Gupter et al. (\citeyear{gupter1998modeling}) have introduced the  exponentiated exponential distribution using the positive power parameter on cumulative distribution function of the standard exponential distribution and utilize it for modeling failure time data. Later on, Nadarajah and Kotz (\citeyear{nadarajah2006exponentiated}) have also proposed other generalized exponentiated type distributions based on some standard distributions such as Weibull, gamma, Fr{\'e}chet, and the gumbel distribution and extensively studied its statistical properties. The exponentiated generalized class of distribution with the addition of two parameters studied by Cordeiro et al. (\citeyear{cordeiro2013exponentiated}) is a more useful model and successfully employed for modeling the lifetime data in many areas of sciences and social sciences. In addition to these models, several alternative models have also been developed in the statistical distribution literature for modeling the lifetime data in the diverse disciplines of sciences. For an excellent overview of the recent work, readers can consult to Cordeiro and de Castro (\citeyear{cordeiro2011new}),  Nadarajah et al. (\citeyear{nadarajah2014new}), Alizadeh et al. (\citeyear{alizadeh2017new}), and Sousa-Ferreira et al. (\citeyear{sousa2023extended}).
	
	In recent decades, much attention has been paid by researchers to studying asymmetric and skewed type data.  Many areas of sciences and social sciences have frequently involved these types of data. However, sometimes these class of distributions do not provide an acceptable fit. Therefore, a distribution having good properties is needed.  Recently,  Shaw and Buckley (\citeyear{shaw2009alchemy}) introduced a new class of transmuted distributions using the continuous distribution as a baseline distribution. Transmutation of the baseline distribution is a powerful tool to construct the skewed probability distribution and has been recently used by several researchers. Aryal and Tsokos (\citeyear{aryal2009transmuted}) have discussed the concept of transmuted extreme value distribution and studied the various mathematical properties for transmuted Gumbel probability distribution. He found in his work that this distribution is more suitable for analyzing the climate data. Moreover, Aryal and Tsokos
	(\citeyear{aryal2011transmuted}) presented a novel extension of the Weibull distribution termed as the transmuted Weibull distribution which captures the good characteristics and is more suitable in the survival analysis.   Khan and King (\citeyear{khan2013transmuted}) considered an extension of the Aryal and Tsokos (\citeyear{aryal2011transmuted}) model and introduced the transmuted modified Weibull distribution.
	They also derived its key properties and utilized the maximum likelihood estimation approach to estimate the unknown parameters. Furthermore, Khan and King (\citeyear{khan2014new}) presented a generalization of the transmuted inverse Weibull distribution which is more applicable in modeling failure criteria and several shapes of aging in reliability analysis.  In addition to these publications, there are a number of other papers in the literature that discuss the fundamental characteristics of various new transmuted distributions, for one glimpse, readers can see Elbatal  (\citeyear{elbatal2013transmuted}), Merovci et al. (\citeyear{merovci2013transmuted}), Tian et al. (\citeyear{tian2014transmuted}), Granzotto and Louzada (\citeyear{granzotto2015transmuted}), Saboor et al. (\citeyear{saboor2016transmuted}), Kemaloglu and Yilmaz (\citeyear{kemaloglu2017transmuted}), Granzotto et al. (\citeyear{granzotto2017cubic}), Bhatti et al. (\citeyear{bhatti2018transmuted}), Tanis et al. (\citeyear{tanics2020transmuted}),  Saracogle and Tanis (\citeyear{saraccouglu2021new}), and Tanis and Saracogle (\citeyear{tanics2023cubic}).
	
	The main motive of this article is to introduce a new class of generalized transmuted lifetime distributions and study its important statistical properties. We abbreviate it as GTLD. The proposed GTLD is a more general family of probability distributions and includes several well-known distributions as the sub-family. The generalized transmuted exponential (GTE), generalized transmuted Rayleigh (GTR), generalized transmuted Weibull (GTW), generalized transmuted modified Weibull (GTMW), generalized transmuted Weibull extension (GTWE), generalized transmuted Burr-type-XII (GTB-XII), generalized transmuted Lomax (GTL), and generalized transmuted pareto type-I (GTP-I) distribution are the important sub-families of the proposed model. These class of distributions have different shapes of hazard rates and are more useful for modeling lifetime data set.
	
	The organization of the article is as follows: Section 2 presents the mathematical construction of the proposed GTLD and discuss its important sub-families. Section 3 presents the some useful expansion of GTLD. Section 4 covers some important statistical properties such as quantile function, moments, moment generating function, stress-strength reliability function, order statistics, R{\'e}nyi and q-entropy,  residual and reversed residual life function, and cumulative information generating function of the proposed GTLD. In Section 5, the maximum likelihood estimators, least squares and weighted least squares estimators, Cram{\'e}r-von Mises estimators, Anderson-Darling estimators, and Right-tail Anderson Darling estimators of the parameters are explored. Section 6 presents an extensive Monte Carlo simulation study for a particular choice of the sub-family of GTLD. Finally, two real data sets are analyzed to show the effectiveness of GTLD in real life modeling and the paper ends with conclusion. 
	

	\section{ Generalized Transmuted Lifetime Distribution}
\begin{figure}[ht] 
		\centering
		\subfloat{\includegraphics[scale=0.43, angle=0]{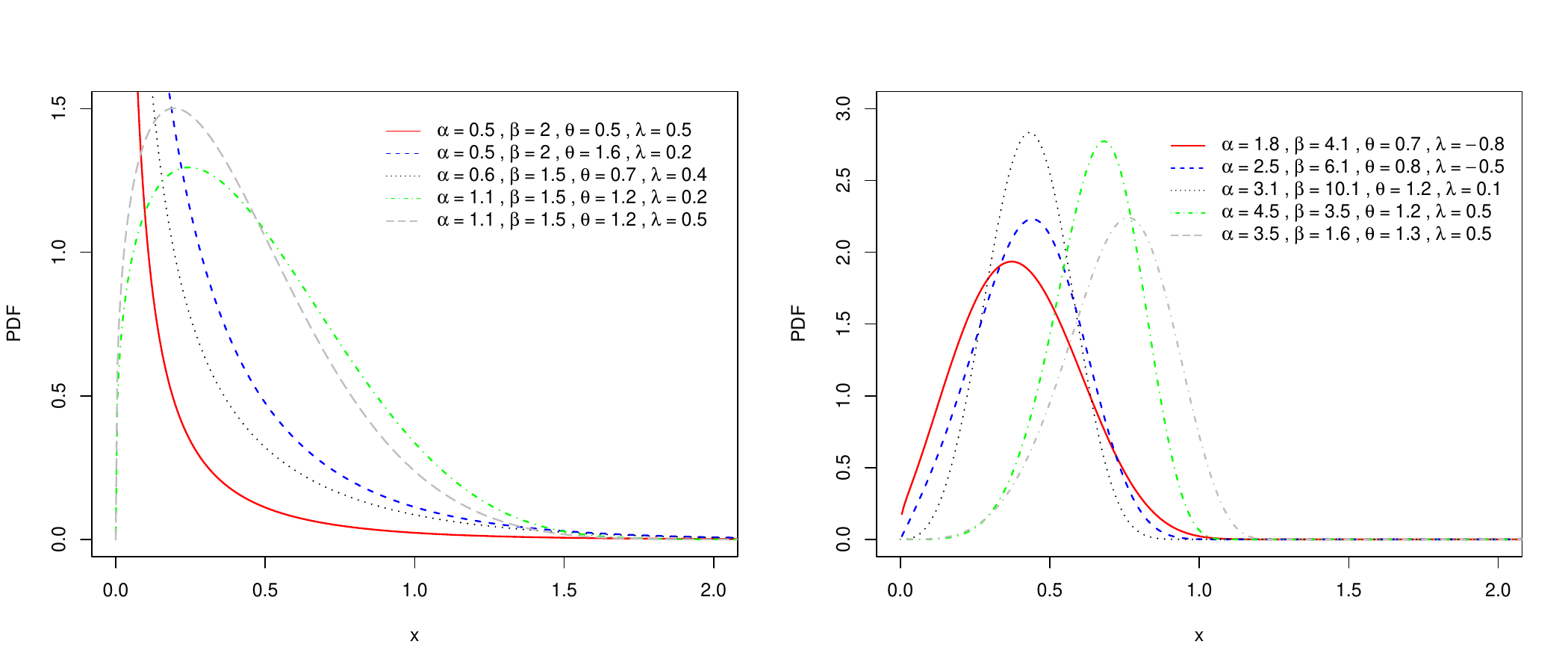}}
		\caption{PDF plots of GTWE distribution.}
		\label{fig.1}
	\end{figure}

\begin{figure}[h] 
		\centering
		\subfloat{\includegraphics[scale=0.43, angle=0]{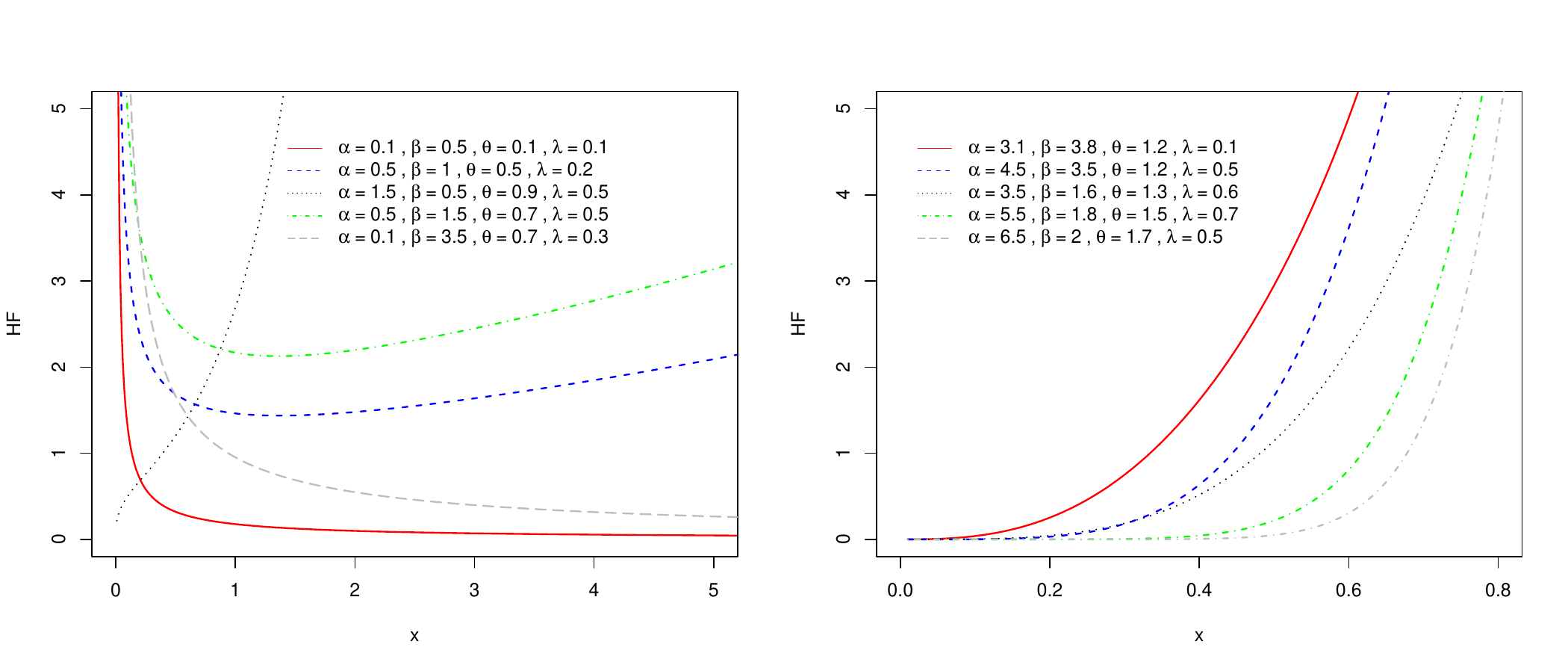}}
		\caption{HF plots of GTWE distribution.}
		\label{fig.2}
	\end{figure}

	\noindent A random variable $U$ is said to have a general lifetime model, which belongs to the shape-scale family if its cumulative distribution function (CDF) is given by
	\begin{equation}\label{baselinecdf}
		H_{U}(x)= 1- \exp{\{-\beta g^{\alpha}(x)\}},~x>0,~ \alpha,\beta>0,
	\end{equation}
	and probability density function (PDF) is given by
	\begin{equation}\label{baselinepdf}
		h_{U}(x)= \alpha \beta g^{\alpha-1}(x) g'(x) \exp{\{-\beta g^{\alpha}(x)\}} ~x>0, ~\alpha,\beta>0,
	\end{equation}
	where $\alpha$ and $\beta$ represents the shape and scale parameter and $g(x)$ is strictly increasing function of $x$ such that $g(0^+)=0$ and $g(x)\rightarrow\infty$ as $x\rightarrow\infty$ (see Maswadah (\citeyear{maswadah2022improved})). We denote it by $U\sim LD_{g}(\alpha,\beta)$. The above defined family includes the most extensively used lifetime distributions by considering the different values of $g^\alpha(x)$.\\
	\noindent Now, we assume that a random variable $T\in [m,n]$ for $-\infty<m<n<\infty$ has density function $v(t)$ and let $Z[H(x;\xi)]$ be  a function of the CDF of a random variable $X$ which satisfies the following conditions:
	\begin{enumerate}[label=(\alph*)]
		\item $Z[H(x;\xi)]\in [m,n]$,
		\item $Z[H(x;\xi)]$ must be monotonically increasing  and can be differentiable,
		\item $Z[H(x;\xi)]\rightarrow m$ as $x\rightarrow-\infty$ and $Z[H(x;\xi)]\rightarrow n$ as $x\rightarrow\infty$.
	\end{enumerate}
	Then, the T-X family of the distribution (see Alzaatreh et al. (\citeyear{alzaatreh2013new})) is defined as
	\begin{equation}
		\label{eq:c1}
		F(x)= \int_{m}^{Z[H(x;\xi)]} v(t) dt,~~ x\in \mathbb{R},
	\end{equation}
	with PDF
	\begin{equation}
		\label{eq:c2}
		f(x)=  \Big\{\frac{\partial }{\partial x} Z[H(x;\xi)]\Big\} v\big \{Z[H(x;\xi)]\big \} ,x\in \mathbb{R}.
	\end{equation}
	For $Z[H(x;\xi)]= [H(x;\xi)]^{\theta}$ and  $v(t)=1+\lambda-2\lambda t;~~  0<t<1,$
	Alizadeh et al. (\citeyear{alizadeh2017generalized}) have presented a new generalized transmuted distribution of the form
	\begin{align} \label{eq:c3}
		\begin{split}
			F(x; \theta, \lambda,\xi)&= \int_{0}^{[H(x;\xi)]^{\theta}} (1+\lambda-2\lambda t) \,dt\\
			&= (1+\lambda) [H(x;\xi)]^{\theta}-\lambda [H(x;\xi)]^{2\theta},
		\end{split}   	
	\end{align}
	where $[H(x;\xi)]^{\theta}$ denotes the baseline family of CDF which depends on parameter vector $\xi$ and $\theta>0,|\lambda|\leq 1$ are two additional parameters.\\
	\noindent We say that a random variable $X$ will have a generalized transmuted class of lifetime distribution if its CDF and PDF is given by
	\begin{equation}\label{cdf}
		F( x; \alpha,\beta,\theta,\lambda )=(1+\lambda) [1- \exp{\{-\beta g^{\alpha}(x)\}}]^{\theta}-\lambda [1- \exp{\{-\beta g^{\alpha}(x)\}}]^{2\theta},
	\end{equation} 
	and
	\begin{equation}\label{pdf}
		\begin{aligned}
			f( x; \alpha,\beta,\theta,\lambda)&= \theta \alpha \beta g^{\alpha-1}(x) g'(x) \exp{\{-\beta g^{\alpha}(x)\}} [1- \exp{\{-\beta g^{\alpha}(x)\}}]^{\theta-1} \\
			& \times\{1+\lambda-2\lambda[1- \exp{\{-\beta g^{\alpha}(x)\}}]^{\theta}\},
		\end{aligned}	
	\end{equation}
	respectively, where $x>0,~\alpha,\beta,\theta> 0,~\text{and}~|\lambda|\leq 1$. We denoted it by GTLD($\alpha,\beta,\theta,\lambda$).
	\noindent From \eqref{cdf}, it can seen that when $\lambda$=0 and $\theta$=1, the GTLD reduces to the lifetime distribution. The hazard function (HF) of the GTLD is given by
	\begin{equation}\label{hazard}
		\begin{aligned}
			&r(x)=
			\frac{\theta \alpha \beta g^{\alpha-1}(x) g'(x) \exp{\{-\beta g^{\alpha}(x)\}} [1- \exp{\{-\beta g^{\alpha}(x)\}}]^{\theta-1} \{1+\lambda-2\lambda[1- \exp{\{-\beta g^{\alpha}(x)\}}]^{\theta}\}}{1-(1+\lambda) [1- \exp{\{-\beta g^{\alpha}(x)\}}]^{\theta}+\lambda [1- \exp{\{-\beta g^{\alpha}(x)\}}]^{2\theta}}.
		\end{aligned}
	\end{equation}
	Figure \ref{fig.1} and Figure \ref{fig.2} shows the plots of the PDF and HF for $g^{\alpha}(x)= e^{x^{\alpha}}-1$, which leads to the generalized transmuted Weibull extension (GTWE) distribution. From these figures, we see that the PDF and HF of the GTWE distribution takes different shapes which suggest the more applicability of the model in diverse ares of research. 
	\subsection{Sub-Families:}
	\noindent Here, we present some important sub-families of GTLD, which encompasses the known distribution families as well as their extensions. 
	\begin{enumerate}[label=(\roman*)]
		\item \textbf{Generalized Transmuted Exponential (GTE) Distribution:}\\
		For $g(x) = x$ and $\alpha=1$, \eqref{cdf} corresponds to the GTE distribution with parameters $\beta$, $\theta$, and $\lambda$ given by
		\begin{equation}\label{GTE}
			F(x)=(1+\lambda) \{1- e^{-\beta x}\}^{\theta}-\lambda \{1- e^{-\beta x}\}^{2\theta},
		\end{equation}
		where $x>0,~\beta,\theta>0$, and $|\lambda|\leq 1$.
		
		\item \textbf{Generalized Transmuted Rayleigh (GTR) Distribution:}\\
		For $g(x) = x^2/2$ and $\alpha=1$, \eqref{cdf} corresponds to the GTR distribution with parameters $\beta$, $\theta$, and $\lambda$. The CDF of GTR is
		\begin{equation}
			F(x)=(1+\lambda) \{1- e^{-\frac{\beta}{2} x^2}\}^{\theta}-\lambda \{1- e^{{-\frac{\beta}{2} x^2}}\}^{2\theta},
		\end{equation}
		where $x>0,~ \beta,\theta>0$, and $|\lambda|\leq 1$.
		
		\item \textbf{Generalized Transmuted Weibull (GTW) Distribution:}\\
		For $g^{\alpha}(x) = x^\alpha$, \eqref{cdf} yields to the GTW distribution with distribution function 
		\begin{equation}\label{GTW}
			F(x)=(1+\lambda) \{1- e^{-\beta x^\alpha}\}^{\theta}-\lambda \{1- e^{-\beta x^\alpha}\}^{2\theta},
		\end{equation}
		where $x>0,~\alpha,\beta,\theta>0$, and $|\lambda|\leq 1$. 
		
		\item \textbf{Generalized Transmuted Modified Weibull (GTMW) Distribution:}\\
		Specially, when $g^{\alpha}(x)=x^\alpha e^{\gamma x}$, \eqref{cdf} corresponds to the GTMW distribution with parameter $\alpha$, $\beta$, $\theta$, $\gamma$,  and $\lambda$. The CDF of GTMW is
		\begin{equation}
			F(x)=(1+\lambda) \{1- e^{-\beta x^\alpha e^{\gamma x}}\}^{\theta}-\lambda \{1- e^{-\beta x^\alpha e^{\gamma x}}\}^{2\theta},
		\end{equation}
		where $x>0,~\alpha,\beta,\theta,\gamma>0$, and $|\lambda|\leq 1$. 
		
		\item \textbf{Generalized Transmuted Weibull Extension (GTWE) Distribution:}\\
		For $g^{\alpha}(x)= e^{x^\alpha}-1$, \eqref{cdf} leads to the GTWE distribution with parameter $\alpha,\beta,\theta$, and $\lambda$. The CDF of GTWE is
		\begin{equation}\label{GTWE}
			F(x)=(1+\lambda) \{1- e^{-\beta[e^{x^\alpha}-1]}\}^{\theta}-\lambda \{1- e^{-\beta[e^{x^\alpha}-1]}\}^{2\theta},
		\end{equation}
		where $x>0,~\alpha,\beta,\theta>0$, and $|\lambda|\leq 1$.	
		\item \textbf{Generalized Transmuted Burr-Type-XII (GTB-XII) Distribution:}\\
		For $g^{\alpha}(x)= \log(1+x^\alpha)$, \eqref{cdf} corresponds to the GTB-XII distribution with parameter $\alpha,\beta,\theta$, and  $\lambda$. The CDF of GTB-XII is
		\begin{equation}
			F(x)=(1+\lambda) \{1-(1+x^\alpha)^{-\beta}\}^{\theta}-\lambda \{1-(1+x^\alpha)^{-\beta}\}^{2\theta},
		\end{equation}
		where $x>0,~\alpha,\beta,\theta>0$, and $|\lambda|\leq 1$.
		
		\item \textbf{Generalized Transmuted Lomax (GTL) Distribution:}\\
		Further, when $g^{\alpha}(x)= \log(1+x/\alpha)$, \eqref{cdf} takes the form of GTL distribution with parameter $\alpha,\beta,\theta$, and  $\lambda$. The CDF of GTL is
		\begin{equation}
			F(x)=(1+\lambda) \{1-(1+x/\alpha)^{-\beta}\}^{\theta}-\lambda \{1-(1+x/\alpha)^{-\beta}\}^{2\theta},
		\end{equation}
		where $x>0,~ \alpha,\beta,\theta>0$, and $|\lambda|\leq 1$.

		\item \textbf{Generalized Transmuted Pareto-Type-I (GTP-I) Distribution:}\\
		Additionally, when $g^{\alpha}(x)= \log(x/\alpha)$, \eqref{cdf} takes the form of GTP-I distribution with parameter $\alpha,\beta,\theta$, and $\lambda$. The CDF of GTP-I is
		\begin{equation}
			F(x)=(1+\lambda) \{1-(x/\alpha)^{-\beta}\}^{\theta}-\lambda \{1-(x/\alpha)^{-\beta}\}^{2\theta},
		\end{equation}
		where $x>0,~ \alpha,\beta,\theta>0$, and $|\lambda|\leq 1$.	
	\end{enumerate}
	\section{Useful Expansion}  Here, we express the PDF \eqref{pdf} of GTLD  as a linear mixture of the densities of the lifetime distribution (LD). Let $X\sim\text{GTLD}(\alpha,\beta,\theta,\lambda)$. Then, the CDF \eqref{cdf} of GTLD is a linear combination of two CDF's of the generalized lifetime distribution (GLD). Similarly, \eqref{pdf} is a linear combination of two PDF's of GLD. By considering the generalized binomial expansion $(1-z)^{\alpha}= \sum_{i=0}^{\infty}(-1)^i \binom{\alpha}{i} z^i,$ \eqref{cdf} can be expressed as
	\begin{equation}\label{cdf2}
		F(x)= \sum_{k=0}^{\infty} c_{k} \exp(-\beta k g^{\alpha}(x)),
	\end{equation}
	where $c_{k}= (-1)^k \left((1+\lambda)\binom{\theta}{k}-\lambda \binom{2\theta}{k}\right)$ with $c_{0}=0$.
	Let  $s_{k+1}= \exp(-\beta (k+1) g^{\alpha}(x))$ (for $k\geq 0$) be the survival function (SF) of LD($\alpha, (k+1)\beta$) with density $h_{k+1}(x)$. Then, the density of GTLD can be written in the following form
	\begin{equation}\label{seriespdf}
		f(x)= \sum_{k=0}^{\infty}d_{k+1} h_{k+1}(x),
	\end{equation}
	where $d_{k+1}=-c_{k+1}$ for $k\geq 0$. Further, we can also use the LD properties to obtain some mathematical properties of the GTLD. \\
	\noindent In addition, we also provide an alternative useful representation for \eqref{pdf}. Using the concept of exponentiated-class, the generalized binomial expansion of the expression $[1- \exp{\{-\beta g^{\alpha}(x)\}}]^{\theta}$ can be expanded into the following infinite series
	\begin{align*}
		[1- \exp{\{-\beta g^{\alpha}(x)\}}]^{\theta}&=\sum_{i=0}^{\infty} (-1)^i \binom{\theta}{i} [\exp{\{-\beta g^{\alpha}(x)\}}]^{i},\\
		&= \sum_{i=0}^{\infty}\sum_{k=0}^{i}(-1)^{i+k} \binom{\theta}{i}\binom{i}{k} [1- \exp{\{-\beta g^{\alpha}(x)\}}]^{k}.
	\end{align*}
	By interchanging the summation indices, the expression is further simplified as follows
	\begin{align*}
		[1- \exp{\{-\beta g^{\alpha}(x)\}}]^{\theta}&= \sum_{k=0}^{\infty}\sum_{i=k}^{\infty}(-1)^{i+k} \binom{\theta}{i}\binom{i}{k} [1- \exp{\{-\beta g^{\alpha}(x)\}}]^{k},\\
		&=\sum_{k=0}^{\infty} S_{k}(\theta)[1- \exp{\{-\beta g^{\alpha}(x)\}}]^{k},
	\end{align*}
	where $S_{k}(\theta)= \sum_{i=k}^{\infty} (-1)^{i+k} \binom{\theta}{i} \binom{i}{k}$. By utilizing the generalized binomial expansion, $F(x)$ can also be written as
	\begin{align*}
		F(x)&= \sum_{k=0}^{\infty} c_{k}  \Pi_{k}(x),\end{align*}
	where
	\begin{equation}\label{ck}
		c_{k}= (1+\lambda) S_{k}(\theta)- \lambda S_{k}(2\theta)
	\end{equation}
	 and $\Pi_{k}(x)=[1- \exp{\{-\beta g^{\alpha}(x)\}}]^{k}$. 
	The function $\Pi_{k}(x)$ represent the CDF of the exponentiated-G distribution. For some more key features of the exponentiated-G distribution, readers can refer to  Gupta and Kundu (\citeyear{gupta1999theory}) and Nadarajah and Kotz (\citeyear{nadarajah2006exponentiated}). Furthermore, the pdf of $X$ can also be written as
	\begin{equation}\label{pdfexp}
		f(x)= \sum_{k=0}^{\infty} c_{k+1} \pi_{k+1}(x),
	\end{equation}
	where $\pi_{k+1}(x)= (k+1)[1- \exp{\{-\beta g^{\alpha}(x)\}}]^{k} \alpha\beta g^{\alpha-1}(x) g'(x) \exp{\{-\beta g^{\alpha}(x)\}} $ (for $k\geq 0$) is the density of exponentiated-G distribution with power parameter $(k)$. This representation underscores that the density function of $\text{GTLD}$ is an infinite linear combination of exponentiated-G density functions. 
\section{Statistical Properties}
\subsection{Quantile Function:} 
A random variable $X$ having CDF $F(x)$, the quantile function (QF) is defined as 
	\begin{equation*}
		Q_{X}(p)=F^{-1}(p)= \inf\{x\in \mathbb{R}^{+}: p\leq F(x)\},~ p\in (0,1)~\text{and}~\mathbb{R}^{+}\in (0,\infty).
	\end{equation*}
	Next, we provide the general expression of QF for GTLD through the following Proposition.
	\begin{proposition}
		Let $X\sim\text{GTLD}(\alpha,\beta,\theta,\lambda)$. Then, the $p$th quantile is given as
		\begin{equation}\label{Prop4.1}
			Q_{X}(p)= g^{-1}\left\{\frac{-1}{\beta} \log (1-A^{1/\theta})\right\}^{\frac{1}{\alpha}},
		\end{equation}
		where $A=\left( \frac{1+\lambda-\sqrt{(1+\lambda)^2-4p\lambda}}{2\lambda}\right)$, $\lambda \neq$ 0.
	\end{proposition}
	\begin{proof}
		For some fixed $p\in (0,1)$, the solution of equation $F(x) = p$, yields the QF. 
		Since, we have $(1+\lambda) [1- \exp{\{-\beta g^{\alpha}(x)\}}]^{\theta}-\lambda [1- \exp{\{-\beta g^{\alpha}(x)\}}]^{2\theta}=p$ and by substituting $y=[1- \exp{\{-\beta g^{\alpha}(x)\}}]^{\theta}$, the quadratic equation becomes $\lambda y^2-(1+\lambda) y+p=0$. On solving the quadratic equation for $y$, we obtain
		\begin{align*}
			y= \frac{(1+\lambda)-\sqrt{(1+\lambda)^2-4p\lambda}}{2\lambda}.
		\end{align*}
		Further, some simple calculation completes the proof.
	\end{proof}
	\begin{example}
		Let $X\sim \text{GTWE}(\alpha,\beta,\theta,\lambda)$ derived in \eqref{GTWE}. Then, the QF is given by
		\begin{equation*}
			Q_{X}(p)= \left[\log\left\{1-\frac{1}{\beta} \log (1-A^{\frac{1}{\theta}})\right\}\right]^{\frac{1}{\alpha}}.
		\end{equation*}
	\end{example}
	\noindent Since, the QF can be used as an alternative to the distribution function and contains various interesting properties which are not shared by the distribution function. Therefore, several new measures have been developed with the help of QF and are widely used in several life testing experiments . Out of them, some important measures are given below:
	\begin{enumerate}[label=(\alph*)]
		\item Measure of location is the median defined by $Q(\frac{1}{2})$;
		\item Moors coefficient of kurtosis (MCK) based on octiles is defined by MCK= $\{Q\left(\frac{7}{8}\right)-Q\left(\frac{5}{8}\right)+Q\left(\frac{3}{8}\right)-Q\left(\frac{1}{8}\right)\}/\{Q\left(\frac{6}{8}\right)-Q\left(\frac{2}{8}\right)\}$;
		\item Bowley's coefficient of skewness (BCS) based on quartiles is defined by 	BCS=$\{Q\left(\frac{3}{4}\right)+Q\left(\frac{1}{4}\right)-2Q\left(\frac{1}{2}\right)\}/\{Q\left(\frac{3}{4}\right)-Q\left(\frac{1}{4}\right)\}$.
	\end{enumerate}
	We can simply calculate all these measures for GTLD using the QF provided in Proposition \ref{Prop4.1}. 
	\subsection{Moments:} The ordinary moments of a distribution are very useful to determine the important characteristics and features such as dispersion, skewness, and kurtosis of a distribution. The $r$th moment of the random variable $X$ is defined as
	\begin{equation*}
		\mu_{r}'=\mathbb{E}[X^r]=\int x^r f_{X}(x)dx.
	\end{equation*}
	\begin{proposition}\label{prop1}
		Let $X\sim \text{GTLD}(\alpha,\beta,\theta,\lambda)$. Then, the $r$th moment is defined as 
		\begin{equation}\label{moment}
			\mu_{r}'= \sum_{i=0}^{\infty} \frac{(-1)^i\theta}{(i+1)} \left[(1+\lambda) \binom{\theta-1}{i}-2\lambda \binom{2\theta-1}{i}\right] \int_{0}^{\infty} \left[g^{-1}\left\{ \frac{y}{(i+1)\beta}\right\}^{1/\alpha}\right]^r e^{-y} dy.
		\end{equation}
	\end{proposition}
	\begin{proof}
		We have 
		\begin{align*}
			\mu_{r}'&= \int_{0}^{\infty} x^r \left[\theta \alpha\beta (1+\lambda) g^{\alpha-1}(x) g'(x) \exp(-\beta g^{\alpha}(x)) \left\{1-\exp(-\beta g^{\alpha}(x))\right\}^{\theta-1}\right]\\
			&- x^r \left[ 2\theta \alpha\beta \lambda g^{\alpha-1}(x) g'(x) \exp(-\beta g^{\alpha}(x)) \left\{1-\exp(-\beta g^{\alpha}(x))\right\}^{2\theta-1}\right] dx.
		\end{align*}
		By the use of generalized binomial expansion, we get
		\begin{align*}
			\mu_{r}'&= \sum_{i=0}^{\infty} (-1)^i \left[(1+\lambda) \binom{\theta-1}{i}-2\lambda \binom{2\theta-1}{i}\right] \int_{0}^{\infty} \theta \alpha \beta x^r g^{\alpha-1}(x) g'(x) \exp\{-(i+1)\beta g^{\alpha}(x)\} dx.
		\end{align*}
		On setting $y=(i+1)\beta g^{\alpha}(x)$ and some simple calculation completes the proof.
	\end{proof}
\noindent Next, we have the following example.
	\begin{example}
		Let $X\sim \text{GTW}(\alpha,\beta,\theta,\lambda)$ derived in \eqref{GTW}. Then
		\begin{equation*}
			\mu_{r}'=\sum_{i=0}^{\infty} (-1)^i \left[(1+\lambda) \binom{\theta-1}{i}-2\lambda \binom{2\theta-1}{i}\right] \frac{\theta}{\beta^{r/\alpha} (i+1)^{r/\alpha+1}} \Gamma\left(\frac{r}{\alpha}+1\right),	
		\end{equation*}
		where $\Gamma(\cdotp)$ is a gamma function defined as $\Gamma(m)=\int_{0}^{\infty} x^{m-1} e^{-x} dx$.
	\end{example}
	\subsection{Incomplete Moments:} The $r$th incomplete moment is defined as
	\begin{equation*}
		\phi_{r}(z)=\int_{0}^{z} x^r f(x) dx.
	\end{equation*}
	\begin{proposition}
		Let $X\sim \text{GTLD}(\alpha,\beta,\theta,\lambda)$. Then, the $r$th incomplete moment of $X$ is given by
		\begin{equation}\label{incompletemoment}
			\phi_{r}(z)= \sum_{i=0}^{\infty} \frac{(-1)^i\theta}{(i+1)} \left[(1+\lambda) \binom{\theta-1}{i}-2\lambda \binom{2\theta-1}{i}\right] \int_{0}^{(i+1)\beta g^{\alpha}(z)} \left[g^{-1}\left\{ \frac{y}{(i+1)\beta}\right\}^{1/\alpha}\right]^r e^{-y} dy.
		\end{equation}
	\end{proposition}
	\begin{proof}
		The proof follows the similar steps as in Proposition \ref{prop1}.
	\end{proof}
	
	\begin{example}
		Let $X\sim \text{GTW}(\alpha,\beta,\theta,\lambda)$ derived in \eqref{GTW}. Then, the $r$th incomplete moment becomes 
		\begin{equation*}
			\phi_{r}(z)=\sum_{i=0}^{\infty} (-1)^i \left[(1+\lambda) \binom{\theta-1}{i}-2\lambda \binom{2\theta-1}{i}\right] \frac{\theta}{\beta^{r/\alpha} (i+1)^{r/\alpha+1}} \gamma\left(\frac{r}{\alpha}+1, (i+1)\beta z^{\alpha}\right),
		\end{equation*}
		where $\gamma(s,x)$ is the lower incomplete gamma function defined as $\gamma(s,x)=\int_{0}^{x} t^{s-1} e^{-t} dt$.
	\end{example}
	\noindent Next, we present the probability weighted moments for the GTLD.
	
	\subsection{Probability Weighted Moments:} The $(r,s)$th probability weighted moments (PWM) of $X$, say $\rho_{r,s}$, is defined as (see Nofal et al. (\citeyear{nofal2017generalized}))
	\begin{equation*}
		\rho_{r,s}= \mathbb{E}[X^r F(x)^s]= \int x^r F(x)^s f(x) dx.
	\end{equation*}
	\begin{proposition}
		Let $X\sim\text{GTLD}(\alpha,\beta,\theta,\lambda)$. Then, 
		\begin{equation*}
			\rho_{r,s}= \sum_{k=0}^{\infty}\frac{\theta}{(l+1)} w_{s,k,l}^* \int_{0}^{\infty} \left[g^{-1}\left\{ \frac{y}{(l+1)\beta}\right\}^{1/\alpha}\right]^r e^{-y} dy,
		\end{equation*} 
		where
		\begin{equation*}
			w_{s,k,l}^{*}= \sum_{l=0}^{\infty}\binom{s}{k} (-1)^{k+l} \lambda^k (1+\lambda)^{s-k} \left[(1+\lambda) \binom{k\theta+s\theta+\theta-1}{l}-2\lambda \binom{k\theta+s\theta+2\theta-1}{l}\right].
		\end{equation*}
	\end{proposition}  
	\begin{proof} The proof is same as that of the Proposition \ref{prop1}.
	\end{proof}
	\begin{example}
		Let $X\sim \text{GTW}(\alpha,\beta,\theta,\lambda)$. Then, the $(r,s)$th PWM is obtained as 
		\begin{equation*}
			\rho_{r,s}=\sum_{k=0}^{\infty} w_{s,k,l}^*	\frac{\theta}{\beta^{r/\alpha} (l+1)^{r/\alpha+1}} \Gamma\left(\frac{r}{\alpha}+1\right).
		\end{equation*}
	\end{example}
	
	\subsection{Moment Generating Function:}
	Let $X\sim\text{GTLD}(\alpha,\beta,\theta,\lambda)$. Then, the moment generating function of the random variable $X$ is given by
	\begin{equation*}
		\begin{aligned}
			&M_{X}(t)= \int_{0}^{\infty} e^{tx} f(x) dx= \sum_{r=0}^{\infty} \frac{t^r}{r!} \int_{0}^{\infty} x^r f(x) dx\\
			&=\sum_{r=0}^{\infty} \sum_{i=0}^{\infty}\frac{t^r}{r!} \frac{(-1)^i\theta}{(i+1)} \left[(1+\lambda) \binom{\theta-1}{i}-2\lambda \binom{2\theta-1}{i}\right] \int_{0}^{\infty} \left[g^{-1}\left\{ \frac{y}{(i+1)\beta}\right\}^{1/\alpha}\right]^r e^{-y} dy.	
		\end{aligned}
	\end{equation*}   
	\begin{example}
		Let $X\sim \text{GTW}(\alpha,\beta,\theta,\lambda$) derived in \eqref{GTW}. Then, we obtain
		\begin{equation*}
			M_{X}(t)=  \sum_{r=0}^{\infty} \sum_{i=0}^{\infty}\frac{t^r}{r!} (-1)^i \left[(1+\lambda) \binom{\theta-1}{i}-2\lambda \binom{2\theta-1}{i}\right] \frac{\theta}{\beta^{r/\alpha} (i+1)^{r/\alpha+1}} \Gamma\left(\frac{r}{\alpha}+1\right).
		\end{equation*}
	\end{example}

	 
	\subsection{Stress-Strength Reliability:} In reliability theory, stress-strength concept is generally used to observe the behavior of a system under stress. Let $X_{1}$ denotes the strength of the system and $X_{2}$ denotes the stress applied on the system. Then, in terms of probability, the stress strength reliability is given by $R=P(X_{1}>X_{2})$.
	\begin{proposition}
		Let $X_{1}\sim\text{GTLD}(\alpha,\beta,\theta,\lambda_{1}$) and $X_{2}\sim\text{GTLD}(\alpha,\beta,\theta,\lambda_{2}$) be independent random variables. Then
			\begin{equation*}
			R=\frac{\lambda_{2}-\lambda_{1}+3}{6}.
		\end{equation*}
	\end{proposition}
	\begin{proof}
		We have
		\begin{equation}\label{Realibility}
			R=\int_{0}^{\infty} f_{1}(x;\alpha,\beta,\theta,\lambda_{1}) F_{2}(x;\alpha,\beta,\theta,\lambda_{2})~dx.
		\end{equation}
		Substituting \eqref{cdf} and \eqref{pdf} in \eqref{Realibility}, we get
		\begin{equation*}
			\begin{aligned}
			   R&= \theta \alpha \beta\int_{0}^{\infty} g^{\alpha-1}(x) g'(x) \exp{\{-\beta g^{\alpha}(x)\}} [1- \exp{\{-\beta g^{\alpha}(x)\}}]^{\theta-1} \\
				&\times\{1+\lambda_{1}-2\lambda_{1}[1- \exp{\{-\beta g^{\alpha}(x)\}}]^{\theta}\} [1- \exp{\{-\beta g^{\alpha}(x)\}}]^{\theta}\\
				&\times \{1+\lambda_{2}-\lambda_{2}[1- \exp{\{-\beta g^{\alpha}(x)\}}]^{\theta}\}~dx.
			\end{aligned}
		\end{equation*}
		Making transformation $y= \exp{\{-\beta g^{\alpha}(x)\}}$, we have $\frac{1}{y}dy= -\alpha\beta g^{\alpha-1}(x)g'(x) dx$. Thus
		\begin{equation}\label{real}
			R= \theta\int_{0}^{1} (1-y)^{2\theta-1} (1+\lambda_{1}-2\lambda_{1} (1-y)^{\theta}) (1+\lambda_{2}-\lambda_{2} (1-y)^{\theta}) dy.
		\end{equation}
		After solving the integral \eqref{real}, we get the desired result.
	\end{proof}
	\subsection{Order Statistics:} Ordered statistics models simply come from the ordered random variables and widely used in several disciplines of research such as in statistical inference, nonparametric statistics, and engineering etc (see Azhad et al. (\citeyear{azhad2021estimation}) and Arshad et al. (\citeyear{arshad2023bayesian})). Let $X_{(1)}, X_{(2)}, \ldots, X_{(n)}$ represent the order statistics of a random sample $X_1, X_2,\ldots, X_n$ of size $n$ taken from  GTLD. Then, the PDF of $r$th order statistics, say $f_{X_{(r)}}(x)$, is defined as
	\begin{equation}\label{order}
		f_{X_{(r)}}(x)= \frac{f(x)}{B(r,n-r+1)}~\left\{F_X(x)\right\}^{r-1}~\left\{1-F_X(x)\right\}^{n-r},
	\end{equation}
	where $B(\cdotp ,\cdotp)$ is the beta function.
	Next, the following theorem shows that the density of $r$th order statistics is a mixture representation of LD densities.
	\begin{theorem}
		The density of $r$th ordered statistics can be expressed as a mixture of LD densities given in \eqref{order1}.
	\end{theorem}
	\begin{proof}
		Since, we have (Formula 0.314 in Gradshteyn and Ryzhik (\citeyear{gradshteyn2014table})) 
		\begin{equation}\label{cj}
			\left(\sum_{s=0}^{\infty} a_{s} u^{s}\right)^j= \sum_{k=0}^{\infty} c_{j,s} u^s,
		\end{equation}
		where\begin{equation*}
			c_{j,0}= a_{0}^j,~ c_{j,s}= (sa_{0})^{-1}\sum_{m=1}^{s} [m(j+1)-s] a_{m} c_{j,s-m}, ~\text{for} ~s\geq 1.
		\end{equation*}
		Thus, we can also express \eqref{order} as
		\begin{equation}\label{or}
			f_{X_{(r)}}(x)= \frac{f(x)}{B(r,n-r+1)}~\sum_{k=0}^{r-1} (-1)^k \left[1-F(x)\right]^{n-r+k}.
		\end{equation}
		Now, we define $a_{s}$=$d_{s+1}$ (for $s\geq 0$) and $u$=$u(x)$=$ \exp\{-\beta g^{\alpha}(x)\}$. Then, the sum defined in \eqref{or} becomes
		\begin{equation*}
			\sum_{k=0}^{r-1}(-1)^k \left(\sum_{s=0}^{\infty} a_{s}u^{s+1}\right)^{n-r+k}= \sum_{k=0}^{r-1}(-1)^k \sum_{s=0}^{\infty} c_{n-r+k,s} u^{n-r+k+s},
		\end{equation*}
		where, we can find the constraints $c_{n-r+k,s}$ by using \eqref{cj}. Based on the mixture from \eqref{seriespdf}, we can write
		\begin{equation*}
			f_{X_{(r)}}(x)=\frac{1}{B(r,n-r+1)}\sum_{k=0}^{r-1}\sum_{j,s=0}^{\infty}(-1)^k a_{j} (j+1)c_{n-r+k,s} \alpha\beta g^{\alpha-1}(x) g'(x) \exp\{-(n-r+k+1+s+j)\beta g^{\alpha}(x)\}.
		\end{equation*} 
		Finally, we get
		\begin{equation}\label{order1}
			f_{X_{(r)}}(x)=\sum_{k=0}^{r-1}\sum_{j,s=0}^{\infty} w_{k,j,s}~ \pi_{n-r+k+1+s+j}(x),	
		\end{equation}
		where $\pi_{n-r+k+1+s+j}(x)$ is the density of LD$(\alpha, \beta(n-r+k+1+s+j))$  and
		\begin{equation*}
			w_{k,j,s}= \frac{(-1)^k (j+1) a_{j} c_{n-r+k,s}}{(n-r+k+1+j+s)}.
		\end{equation*}
	\end{proof}
	
	\subsection{Entropies:}
	\noindent In statistics, generally, entropies are used to measure the variation or uncertainty associated with a random variable $X$. Several types of entropies have been discussed in literature such as R{\'e}nyi, Shannon, and q-entropy, etc. Here, we have mainly focused on the two well-known entropy measures; namely R{\'e}nyi and q-entropy. For a random variable $X$ having PDF $f(x)$, the R{\'e}nyi and q-entropy, respectively are defined as
	\begin{equation}\label{renyi}
		I_{\rho}(X)= \frac{1}{(1-\rho)}\log\left(\int_{0}^{\infty} f^{\rho}(x)~dx \right),~\rho>0~ \text{and}~ \rho\neq 1
	\end{equation}
	and
	\begin{equation}
		H_{q}(X)= \frac{1}{q-1} \log\left\{1- \int_{0}^{\infty} f^{q}(x) dx\right\}, ~q>0~ \text{and}~ q\neq 1.
	\end{equation}
	The following Propositions present the R{\'e}nyi and q-entropy for the GTLD.  
	\begin{proposition}\label{Prop4.4}
		Let $X$ be a random variable having $\text{GTLD}(\alpha,\beta,\theta,\lambda)$. Then, the R{\'e}nyi entropy is defined as
		\begin{equation*}
			I_{\rho}(X)= \frac{1}{1-\rho} \log\left[w_{j,k}
			\int_{0}^{\infty} e^{-(k+\rho) \beta g^{\alpha}(x)} \{\alpha\beta g^{\alpha-1}(x)g'(x)\}^\rho~dx\right],
		\end{equation*}
		where $w_{j,k}= \theta^\rho \sum_{k=0}^{\infty}\sum_{j=0}^{\infty} (-1)^{j+k}\binom{\rho}{j}\binom{j\theta+\rho(\theta-1)}{k} (2\lambda)^{j} (1+\lambda)^{\rho-j}.$ 
	\end{proposition}
	\begin{proof}
		From \eqref{pdf}, we have	
		\begin{equation*}
			f^\rho(x)= \left[\theta \alpha \beta g^{\alpha-1}(x) g'(x) e^{\{-\beta g^{\alpha}(x)\}} [1- e^{\{-\beta g^{\alpha}(x)\}}]^{\theta-1}\{1+\lambda-2\lambda[1- e^{\{-\beta g^{\alpha}(x)\}}]^{\theta}\}\right]^{\rho}.
		\end{equation*}
		With the help of generalized binomial expansion, we get
		\begin{equation}\label{11}
			f^\rho(x)= w_{j,k}~ e^{-(k+\rho) \beta g^{\alpha}(x)} \{\alpha\beta g^{\alpha-1}(x)g'(x)\}^\rho.
		\end{equation}
		On putting \eqref{11} in \eqref{renyi}, we get the desired result.
	\end{proof} 
	\begin{proposition}
		Considering a random variable $X$ having $\text{GTLD}(\alpha,\beta,\theta,\lambda)$. Then, the q-entropy is defined as
		\begin{equation*}
			H_{q}(X)=\frac{1}{q-1} \log\left\{1- \left[w_{j,k}^{*}
			\int_{0}^{\infty} e^{-(k+q) \beta g^{\alpha}(x)} \{\alpha\beta g^{\alpha-1}(x)g'(x)\}^q~dx\right]\right\},
		\end{equation*}
		where $q>0$, $q\neq 1$ and $w_{j,k}^{*}= \theta^q \sum_{k=0}^{\infty}\sum_{j=0}^{\infty} (-1)^{j+k}\binom{q}{j}\binom{j\theta+q(\theta-1)}{k} (2\lambda)^{j} (1+\lambda)^{q-j}.$
	\end{proposition} 
	\begin{example}\label{exqentropy}
		Let $X\sim\text{GTWE}(\alpha,\beta,\theta,\lambda)$. Then the R{\'e}nyi and q-entropy are given by
		\begin{equation*}
			I_{\rho}(X)= \frac{1}{1-\rho} \log\left[w_{j,k}
			\int_{0}^{\infty} e^{-(k+\rho) \beta \left\{e^{x^\alpha}-1\right\}} \{\alpha\beta x^{\alpha-1} e^{x^\alpha}\}^\rho~dx\right]
		\end{equation*}	
		and
		\begin{equation*}
			H_{q}(X)= \frac{1}{q-1} \log\left\{1-\left[w_{j,k}^*
			\int_{0}^{\infty} e^{-(k+\rho) \beta \left\{e^{x^\alpha}-1\right\}} \{\alpha\beta x^{\alpha-1} e^{x^\alpha}\}^q~dx\right]\right\},
		\end{equation*}
	\end{example}
	\noindent respectively. Since, the integral defined in the above example is quite difficult. Hence, we can easily find these measures using the numerical approximation. 
	Table \ref{re} and Table \ref{qe} presents the R{\'e}nyi and q entropy measure for GTWE distribution with some fixed values of parameter $\zeta=(\alpha,\beta,\theta,\lambda)$.
	\begin{table}[ht]
	\caption{ R{\'e}nyi entropy for GTWE distribution with $\zeta=(0.5,2.0,0.5,0.5)$.} 
	\centering 
	\footnotesize
	\begin{tabular}{ccccccccc} 
		\toprule
		$\rho$ &$2$& $3$& $4$&$5$&$6$&$7$&$8$ &$9$\\
		\hline
		\toprule
		$I_{\rho}(X)$& -0.228307& -0.993325& -1.327645& -1.52248& -1.651964& -1.745015& -1.815485&-1.870913\\
		\hline
		
	\end{tabular}\label{re}
	\end{table}	
	
	\begin{table}[ht]
	\caption{ q-entropy for GTWE distribution with $\zeta=(0.1,0.1,0.5,0.5)$.} 
	\centering 
	\footnotesize
	\begin{tabular}{ccccccccc} 
		\toprule
		$q$ &$2$& $3$& $4$&$5$&$6$&$7$&$8$ &$9$\\
		\hline
		\toprule
		$H_{q}(X)$& -0.009695& -0.002006& -0.000772 &-0.00038& -0.000214& -0.000131 &-8.4e-05& -5.7e-05\\
		\hline
	\end{tabular}\label{qe}
	\end{table}	
	\subsection{ Residual Life Function:} 
	The $n$th moment of the residual life of random variable $X$ is defined as (see Merovci et al. (\citeyear{merovci2017exponentiated}))
	\begin{align*}
		\begin{split}
			m_{n}(t)&= E[(X-t)^n| X>t],~ n=1,2,\ldots.\\
			&=\frac{1}{1-F(t)} \int_{t}^{\infty} (x-t)^n dF(x).
		\end{split}
	\end{align*}
	
	\begin{proposition}\label{prop4.7}
		Let $X\sim\text{GTLD}(\alpha,\beta,\theta,\lambda)$. Then, we have
		\begin{equation}\label{RL}
			m_{n}(t)= \frac{1}{1-F(t)} \sum_{r=0}^{n} \binom{n}{r}(-t)^{n-r} \sum_{k=0}^{\infty} c_{k+1}\int_{t}^{\infty} x^r \pi_{k+1}(x).
		\end{equation} 
	\end{proposition}
	\begin{proof}
		By the use of \eqref{pdfexp} and using generalized binomial expansion, we get the Proposition.
	\end{proof}
	\begin{example}
		Let $X\sim\text{GTW}(\alpha,\beta,\theta,\lambda)$. Then
		\begin{equation*}
			m_{n}(t)= \frac{1}{1-F(t)} \sum_{r=0}^{n}\sum_{k=0}^{\infty}\sum_{m=0}^{\infty}(-1)^{m} \binom{n}{r}\binom{k}{m}(-t)^{n-r}c_{k+1}(k+1) \frac{\Gamma\left(\frac{r}{\alpha}+1, \beta (m+1)t^{\alpha}\right)}{\beta^{r/\alpha} (m+1)^{r/\alpha+1}},
		\end{equation*}
	\end{example}
	\noindent where $\Gamma(\cdotp,\cdotp)$ is upper incomplete gamma function and $c_{k+1}$ can be evaluated from \eqref{ck}.\\
	Another more interesting concept in reliability theory is the mean residual life (MRL) which denotes the expected additional life age for a component or system while it has already lived at age $t$ and defined by $m_{1}(t)= E[(X-t)| X>t]$. This function is more useful in several life testing experiments. We can simply obtain MRL for GTLD by taking  $n$=1 in \eqref{RL}.

	\subsection{Reversed Residual Life Function:} The $n$th moment of the reversed residual life is defined as (see Merovci et al. (\citeyear{merovci2017exponentiated}))
	\begin{align*}
		\begin{split} 
			M_{n}(t)&= E[(t-X)^n| X\leq t], n=1,2,\ldots\\
			&= \frac{1}{F(t)} \int_{0}^{t} (t-x)^n dF(x).
		\end{split} 
	\end{align*}
	
	\begin{proposition}
		Let $X\sim\text{GTLD}(\alpha,\beta,\theta,\lambda)$. Then, we have 
		\begin{equation}\label{MIT}
			M_{n}(t)= \frac{1}{F(t)} \sum_{r=0}^{n}(-1)^r \binom{n}{r}(t)^{n-r} \sum_{k=0}^{\infty} c_{k+1}\int_{0}^{t} x^r \pi_{k+1}(x).
		\end{equation} 
	\end{proposition}
	\begin{example}
		Let $X\sim \text{GTW}(\alpha,\beta,\theta,\lambda)$. Then
		\begin{equation*}
			M_{n}(t)= \frac{1}{F(t)} \sum_{r=0}^{n}\sum_{k=0}^{\infty}\sum_{m=0}^{\infty}(-1)^{m+r} \binom{n}{r}\binom{k}{m}(t)^{n-r}c_{k+1}(k+1) \frac{\gamma\left(\frac{r}{\alpha}+1, \beta (m+1)t^{\alpha}\right)}{\beta^{r/\alpha} (m+1)^{r/\alpha+1}},
		\end{equation*}
	\end{example}
	\noindent where $\gamma(\cdotp,\cdotp)$ is the lower incomplete gamma function and $c_{k+1}$ can be evaluated from \eqref{ck}. Further, same as MRL function, mean reversed residual life (MRRL) function or mean waiting time also play an important role in reliability theory. This function denotes the waiting time elapsed of a component or a system while the failure of a component has already occurred in $(0, t )$ and defined by $M_{1}(t)= E[(t-X)| X\leq t]$. 
	
	\subsection{Cumulative Information Generating Function:}
	\noindent Capaldo et al. (\citeyear{capaldo2023cumulative}) have introduced a new generating function which allows to measure the cumulative information coming both from the CDF and the survival function (SF). Let $X$ be a random variable having CDF $F(x)$ and  SF $\overline{F}(x)$. Then, the cumulative information generating function (CIGF) of $X$, denoted by $G_X$, is defined as
	\begin{equation*}
		\begin{aligned}
			(m,n)&\rightarrow G_{X}(m,n)= \int_{l}^{r} [F(x)]^m [\overline{F}(x)]^n dx
		\end{aligned}
	\end{equation*} 
	where $(m, n) \in D_X \subseteq \mathbb{R}^2$; $D_X$ is the set of pairs $(m, n)$ for which $G_X(m, n)$ is finite.\\
    For GTLD we have the following result.
	\begin{proposition}\label{Prop4.6}
		Let $X\sim\text{GTLD}(\alpha,\beta,\theta,\lambda)$. Then, the CIGF of $X$ is expressed as
		\begin{equation*}
			G_X(m,n)= \sum_{i=0}^{\infty}\sum_{j=0}^{\infty} (-1)^{i+j} \binom{n}{i}\binom{m+i}{j}\lambda^j(1+\lambda)^{m+i-j}\int_{0}^{\infty} \left\{1-\exp{(-\beta g^{\alpha}(x))}\right\}^{\theta (j+m+i)}dx.
		\end{equation*}
	\end{proposition}
	\noindent Based on the CIGF,  Capaldo et al. (\citeyear{capaldo2023cumulative}) have also introduced two another measure which are the marginal versions of the CIGF known as the cumulative information generating measure (CIGM) given by
	\begin{equation*}
		H_{X}(m)\equiv G_{X}(m,0)= \int_{l}^{r}[F(x)]^m dx, ~ \forall (m,0)\in D_{X}
	\end{equation*}
	and the cumulative residual information generating measure  (CRIGM) given by
	\begin{equation*}
		K_{X}(n)\equiv G_{X}(0,n)= \int_{l}^{r}[\overline{F}(x)]^n dx, ~ \forall (0,n)\in D_{X}.
	\end{equation*}
	The following Proposition gives the general expression of these measures for GTLD. 
	\begin{proposition}
		Let $X\sim\text{GTLD}(\alpha,\beta,\theta,\lambda)$. Then, the CIGM and CRIGM are respectively
		\begin{equation*}
			H_{X}(m)\equiv G_{X}(m,0)= \sum_{j=0}^{\infty} (-1)^{j} \lambda^j (1+\lambda)^{m-j} \binom{m}{j}  \int_{0}^{\infty} \left\{1-\exp{(-\beta g^{\alpha}(x))}\right\}^{\theta (j+m)}dx
		\end{equation*}
		and 
		\begin{equation*}
			K_{X}(n)\equiv G_{X}(0,n)= \sum_{i=0}^{\infty} \sum_{j=0}^{\infty}(-1)^{i+j} \binom{n}{i}\binom{i}{j} \lambda^j (1+\lambda)^{i-j} \int_{0}^{\infty} \left\{1-\exp{(-\beta g^{\alpha}(x))}\right\}^{\theta (j+i)}dx.
		\end{equation*}
	\end{proposition}

	\section{Parameter Estimation}
   This section examines the estimation of unknown parameters of GTLD. Several approaches of point estimation such as maximum likelihood (ML), ordinary least square (OLS), weighted least square (WLS), Cram{\'e}r-von Mises (CvM), Anderson Darling (AD) and Right-tail Anderson Darling (RTAD)  are applied to calculate the estimators for unknown parameters of the proposed GTLD.
	\subsection{Maximum Likelihood (ML) Estimation:} Several approaches have been introduced in literature to estimate the unknown parameters in which maximum likelihood technique is generally used for estimation purpose. The log-likelihood function based on the sample $x_1, x_2, \ldots, x_n$ of size $n$ taken from GTLD$(\alpha, \beta, \theta, \lambda)$ by using the density function defined in \eqref{pdf}, takes the form
	\begin{equation}\label{likelihood}
		\begin{aligned}
			l(\zeta|\textbf{x})&= n\left(\log \theta+\log \alpha+\log\beta\right)+(\alpha-1)\sum_{i=1}^{n}\log g(x_{i})+ \sum_{i=1}^{n} \log g'(x_{i})- \beta \sum_{i=1}^{n} g^{\alpha}(x_{i}) \\
			&+(\theta-1)\sum_{i=1}^{n}\log\left\{1-e^{\left( -\beta g^{\alpha}(x_{i})\right)}\right\}+\sum_{i=1}^{n} \log \left\{1+\lambda-2\lambda \left\{1-e^{\left( -\beta g^{\alpha}(x_{i})\right)}\right\}^{\theta} \right\}.
		\end{aligned}
	\end{equation}
	where $\textbf{x}=(x_{1},x_{2},\ldots,x_{n})$ and $\zeta$=($\alpha$, $\beta$, $\theta$, $\lambda$). The equation obtained in \eqref{likelihood} is of general nature and can be used for several sub-family of GTLD by substituting the $g^{\alpha}(x)$. We can easily obtain the maximum likelihood estimates of parameters on partially differentiating \eqref{likelihood} with respect to $\zeta$. The normal equations obtained through \eqref{likelihood} for any sub-family of GTLD are of non-linear nature. Therefore, we apply the Broyden-Flecther-Goldfarb-Shanno (BFGS) technique to solve the non-linear equations. This technique can be easily applied using optim() function in `stats' package of R-programming library. The following theorems shows the existence and uniqueness of the ML estimates for the parameter $\theta$.
	
	
	\begin{theorem}
		Let the parameters $\alpha$, $\beta$, and $\lambda$ are known and $\lambda\in (-1,0)$. Then, there exist at least one MLE for the parameter $\theta$ which belongs to the interval
		$\left[\frac{n}{-2\sum_{i=1}^{n} \log (y_{i})}, \frac{n}{-\sum_{i=1}^{n} \log (y_{i})} \right]$.
	\end{theorem}
	\begin{proof}
		Let $y_{i}= 1-\exp(-\beta g^{\alpha}(x_{i}))$ and on partially differentiating \eqref{likelihood} with respect to parameter $\theta$, we get  \\
		\begin{equation}\label{partialtheta}
			\frac{\partial}{\partial \theta} l(\zeta|{\bf{x}})= \frac{n}{\theta} +\sum_{i=1}^{n} \log y_{i}-2\lambda\sum_{i=1}^{n}	\frac{y_{i}^{\theta} \log y_{i} }{1+\lambda-2\lambda y_{i}^{\theta}}.	
		\end{equation}
		Since $0<y_{i}<1$ and $-1<\lambda<0$. We have 	$-2\lambda\sum_{i=1}^{n} \frac{y_{i}^{\theta} \log y_{i}}{1+\lambda-2\lambda y_{i}^{\theta}}<0,$ implies that
		\begin{equation*}
			\frac{\partial}{\partial \theta} l(\zeta|{\bf{x}}) <\frac{n}{\theta} +\sum_{i=1}^{n} \log y_{i}.
		\end{equation*}
		Thus, for $\theta> \frac{n}{-\sum_{i=1}^{n}\log y_{i}}$, we obtain that $\frac{\partial}{\partial \theta} l(\zeta|{\bf{x}})$ is negative.\\
		Again, the quantity $0< \frac{-2\lambda y_{i}^{\theta}}{1+\lambda-2\lambda y_{i}^{\theta}}<1$ implies that 
		$\frac{-2\lambda y_{i}^{\theta}}{1+\lambda-2\lambda y_{i}^{\theta}} >\log y_{i},~ \text{and}$
		\begin{equation*}
			\frac{\partial}{\partial \theta} l(\zeta|{\bf{x}})> \frac{n}{\theta} + 2\sum_{i=1}^{n} \log y_{i}.
		\end{equation*}
		Finally, for $\theta< \frac{n}{-2 \sum_{i=1}^{n} \log y_{i}} $, we obtain that $\frac{\partial}{\partial \theta} l(\zeta|{\bf{x}})$ is positive.\\ 
		Therefore, the proof follows from the continuity of  function $\frac{\partial}{\partial \theta} l(\zeta|{\bf{x}})$.
	\end{proof}
	
	
	\begin{theorem}
		Let the parameters $\alpha$, $\beta$, and $\lambda$ are known and  $\lambda\in (0,1)$. Then, there exist an unique MLE for the parameter $\theta$.
	\end{theorem}
	\begin{proof}
		
		From \eqref{partialtheta}, we have
		\begin{equation*}
			\frac{\partial^2}{\partial \theta^2} l(\zeta|{\bf{x}})= -\frac{n}{\theta^2} -2\lambda (1+\lambda)\sum_{i=1}^{n} \frac{ y_{i}^{\theta}(\log y_{i})^2}{(1+\lambda-2\lambda y_{i}^\theta)^2}. 
		\end{equation*}
		Since $\lambda\in (0,1)$ implies that  $\frac{\partial^2}{\partial \theta^2} l(\zeta|{\bf{x}})<0$, which means that $\frac{\partial}{\partial \theta} l(\zeta|{\bf{x}})$  is a decreasing function. 
		Also, we have that $\lim_{\theta\rightarrow 0}\frac{\partial}{\partial \theta} l(\zeta|{\bf{x}})=\infty $ and $\lim_{\theta\rightarrow\infty}\frac{\partial}{\partial \theta} l(\zeta|{\bf{x}})= \sum_{i=1}^{n} \log (y_{i})<0 $, which proves the uniqueness of the MLE of $\theta$.                                                                                     
	\end{proof}
	\noindent On a same line, the existence and uniqueness of other parameters can be also established.
	
	
	\subsection{Ordinary Least Squares (OLS) Estimation:}
	Let $X_{(1)}, X_{(2)}, \ldots, X_{(n)}$ denote the order statistics of a random sample $X_1, X_2,\ldots, X_n$ of size $n$ taken from  GTLD. Then, the OLS estimates of the unknown parameters $\alpha$, $\beta$, $\theta$, and $\lambda$ are obtained by minimizing
	\begin{equation*}\label{OLS}
		Z(\alpha,\beta,\theta,\lambda)=\sum_{i=1}^{n} \left\{F(X_{(i)})-c_{(i,n)}^{[1]}\right\}^2,
	\end{equation*}
	where, $F(X_{(i)})$ denotes the empirical cumulative distribution function
	(ecdf) and $c_{(i,n)}^{[1]}=\frac{i}{n+1}; i= 1, 2, \ldots, n $ is the mean of $F(X_{(i)})$.

	
	\subsection{Weighted Least Squares (WLS) Estimation:} The WLS estimator follows a similar
	procedure to the OLS estimator, where the objective is to minimize the weighted sum of squares differences. The WLS estimates of the unknown parameters $\alpha$, $\beta$, $\theta$, and $\lambda$ are obtained by minimizing the following equation
	\begin{equation*}\label{WLS}
		W(\alpha,\beta,\theta,\lambda)= \sum_{i=1}^{n} c_{i,n}^{[2]} \left[F(X_{(i)})- c_{(i,n)}^{[1]}\right]^2,
	\end{equation*}
	where, $c_{(i,n)}^{[2]}=\frac{(n+1)^2(n+2)}{i(n-i+1)}; i= 1, 2, \ldots, n $ is the inverse of the variance of $F(X_{(i)})$.
	
	\subsection{Cram{\'e}r-von Mises (CvM) Estimation:}
	The CvM estimates of unknown parameters are obtained by minimizing the following equation with respect to unknown parameters $\alpha$, $\beta$, $\theta$, and $\lambda$
	\begin{equation*}\label{CVM}
		C(\alpha,\beta,\theta,\lambda)=\frac{1}{12n}+\sum_{i=1}^{n}\left(F(X_{(i)})-\frac{2i-1}{2n}\right)^2.
	\end{equation*}
	
	
	\subsection{Anderson-Darling (AD) Estimation:}
	The AD estimates of unknown  parameters $\alpha$, $\beta$, $\theta$, and $\lambda$ are obtained  by minimizing the following expression 
	\begin{equation*}\label{AD}
		A(\alpha,\beta,\theta,\lambda)= -n-\frac{1}{n}\sum_{i=1}^{n}(2i-1)\left[ \log F(X_{(i)})+\log\overline{F}(X_{(n+1-i)})\right],
	\end{equation*}
	where, $\overline{F}(x)=1-F(x)$  denotes the SF of GTLD. 
	
	\subsection{The Right-tail Anderson Darling (RTAD) Estimation:}
	The RTAD estimates are obtained by minimizing the following equation with respect to the unknown parameters $\alpha$, $\beta$, $\theta$, and $\lambda$ 
	\begin{equation*}\label{RTADE}
		R(\alpha,\beta,\theta,\lambda)= \frac{n}{2}-2\sum_{i=1}^{n} F(X_{(i)})-\frac{1}{n} \sum_{i=1}^{n} (2i-1) \left[\log \overline{F}(X_{(n+1-i)})\right].
	\end{equation*}
	For more details of the above discussed estimation approaches, readers can refer to Dey et al.   (\citeyear{dey2018kumaraswamy}), Dey et al. (\citeyear{dey2018statistical}), and Arshad et al. (\citeyear{arshad2022record}).
	
	
	\section{Simulation Study}
	 In this section, a well-organized Monte Carlo simulation study is carried out to asses the performance of ML, OLS, WLS, CvM, AD, and  RTAD estimators of unknown parameters of the proposed GTLD$(\alpha, \beta, \theta,\lambda)$. For analysis purpose, we consider a sub-family of GTLD by taking $g^{\alpha}(x)= e^{x^\alpha}-1$ which reduces to GTWE distribution derived in \eqref{GTWE}. A well known Broyden-Fletcher-Goldfarb-Shanno (BFGS) technique introduced by Broyden (\citeyear{broyden1970convergence}), Fletcher (\citeyear{fletcher1970new}), Goldfarb (\citeyear{goldfarb1970family}), and Shanno (\citeyear{shanno1970conditioning}) is used to obtain the estimates of population parameters $\alpha, \beta,\theta,$ and $\lambda$. This optimization technique is easily available in R (version 4.3.3) programming library.  Using inverse CDF method, we generate the data varying from small to large sample sizes i.e. $n$= $ 50, 100, 150, 200, 300$, and $400$ with some fix configurations of parameter values $\zeta$=($\alpha, \beta, \theta, \lambda$).  The performance of all the estimators are observed by using the criteria of absolute biases and the mean squared errors (MSEs). The mathematical formulae of these measures are as follows; absolute bias= $N^{-1} \sum_{i=1}^{N} |(\hat{\zeta}_i-\zeta)|$, mean squared errors (MSEs)= $N^{-1}\sum_{i=1}^{N} (\hat{\zeta}_i-\zeta)^2$, where $\zeta$ is the true value of the parameter, $\hat{\zeta}$ is the estimated value of the parameter $\zeta$ for the $i$th repeated sample, and $N$ be the number of repeated samples.
	The absolute biases and the mean squared errors (MSEs) of considered estimators are calculated corresponding to each sample. This process is replicated $500$ times, and average value of absolute bias and MSEs are reported in Table \ref{T1} to Table \ref{T3}. From Table \ref{T1}-\ref{T3}, we can observe that for some fix configuration of parameters, i.e., $\zeta$=(2.5, 3.0, 0.5, 0.2), $\zeta$=(1.5, 2.0, 0.9, 0.8), and $\zeta$=(2.0, 2.5, 0.7, 0.5); the absolute biases and the MSEs decreases as we increase the sample size.  Moreover, we also note the exceptions for parameter $\alpha$ and $\beta$ in Table \ref{T1} and Table \ref{T3}, as well as, for parameter $\alpha$ in Table \ref{T2} when we use the CvM method.  
	Overall, the estimators demonstrate an improved performance in terms of both the considered measures with larger sample sizes.  Apart from that, we also observe that the CvM technique works worse in all the considered setting of parameter. Based on the results obtained in simulation study, we recommend the ML, WLS, AD, and RTAD estimation techniques for the estimation of parameters of the considered sub-family of GTLD.
	\begin{table}[h]
		\caption{Absolute biases and MSEs of the estimators for settings of parameters $\zeta$=(2.5, 3.0, 0.5, 0.2).}  
		\centering 
		\footnotesize
		\begin{tabular}{ccccccccccc} 
			\hline
			\toprule 
			$N=500$&$\downarrow$Sample& \multicolumn{4}{c}{Absolute Bias} &&\multicolumn{2}{c}{\hspace{2.1cm}MSE} &&\\
			\cmidrule{3-6} \cmidrule{8-11}  
			Estimators &(n)& $\hat{\alpha}$& $\hat{\beta}$ &$\hat{\theta}$&$\hat{\lambda}$&&$\hat{\alpha}$& $\hat{\beta}$ &$\hat{\theta}$&$\hat{\lambda}$  \\
			\hline	
			\toprule 
			&&&&&&&&&&\\
			ML&50&1.06900& 0.98419& 0.23732& 0.33998&& 1.66527&  1.84613& 0.15532& 0.15618\\
			&100&0.92468& 0.79680& 0.16357& 0.32976&& 1.22863&  1.23454& 0.04488& 0.14615\\
			&150&0.80716& 0.61172 &0.12924 &0.35603&& 0.97735 &0.62471& 0.02305& 0.16149\\
			&200&0.76561& 0.50938& 0.11847& 0.32530&& 0.89849 &0.43978& 0.01953& 0.13691\\
			&300&0.68259& 0.39266& 0.10135& 0.30068&& 0.76864& 0.25828& 0.01487& 0.12130\\
			&400&0.49661& 0.26532& 0.07468& 0.24026&& 0.49251 &0.12359& 0.00940& 0.08365\\
			
			&&&&&&&&&&\\
			\hline
			&&&&&&&&&&\\
			OLS&50&2.27208& 2.24970& 0.83155& 0.30068&& 9.37800& 21.84073& 5.51961& 0.13895\\
			&100&1.87174& 1.74631& 0.35958& 0.26307&& 6.65172& 11.95662& 0.37070& 0.10762\\
			&150&1.39442& 1.20046 &0.23436 &0.25524 &&4.00471& 5.02845 &0.14657 &0.09951\\
			&200&1.19043& 0.87403& 0.17763& 0.25336&& 3.28721& 2.03997& 0.06310& 0.09202\\
			&300&0.86044& 0.56817& 0.12359& 0.23590&& 1.95893& 0.61100& 0.02497 &0.07946\\
			&400&0.55665& 0.30541& 0.08745& 0.18784&& 0.78282& 0.16649& 0.01258& 0.04699\\

			&&&&&&&&&&\\
			\hline
			&&&&&&&&&&\\
			WLS&50&1.92467& 1.78464& 0.53200& 0.60821&& 7.15801&  1.61775& 1.35731& 0.13895\\
			&100&1.52066& 1.28522& 0.25609& 0.31963&& 4.90048&  0.23737& 0.14424& 0.10762\\
			&150&1.10357& 0.85408 &0.16679& 0.26899 &&3.02167& 0.12297& 0.05539& 0.09951\\
			&200&0.89269& 0.63325& 0.13937& 0.25258&& 1.99180& 0.09306& 0.03330& 0.09202\\
			&300&0.60249& 0.39566& 0.09479& 0.25383&& 1.04372 &0.07710& 0.01518& 0.07946\\
			&400&0.36202& 0.19316& 0.06518& 0.28258 &&0.30470& 0.08652& 0.00698& 0.04699\\
		
			&&&&&&&&&&\\
			\hline
			&&&&&&&&&&\\
			CvM&50&0.29260& 0.18823& 0.06581& 0.10477&& 0.08701 & 0.03914& 0.00562 &0.01104\\
			&100&0.30197& 0.19241& 0.06657& 0.10179&& 0.09142 & 0.03942& 0.00532 &0.01037\\
			&150&0.30405 &0.19357& 0.06656 &0.10105&& 0.09252& 0.04024& 0.00510 &0.01021\\
			&200&0.30512& 0.19306& 0.06736& 0.10067&& 0.09313& 0.03821 &0.00503 &0.01014\\
			&300&0.30569& 0.19739& 0.06819& 0.10044&& 0.09346 &0.03922& 0.00491& 0.01009\\
			&400&0.30568& 0.19983& 0.06936& 0.10034&& 0.09345& 0.03993 &0.00491& 0.01007\\
		
			&&&&&&&&&&\\
			\hline
			&&&&&&&&&&\\
			AD&50&1.27747& 1.36121 &0.42101& 0.66662&& 2.07828&  2.19135& 0.60813& 0.49107\\
			&100&1.22846& 1.47607& 0.25846& 0.73435 &&1.89970&  2.41233 &0.15318& 0.56040\\
			&150&1.18576& 1.51542& 0.19884 &0.75542 &&1.76883 &2.44775& 0.06787 &0.57723\\
			&200&1.19629& 1.54922& 0.17706 &0.76302 &&1.77899& 2.50320& 0.04620& 0.58719\\
			&300&1.19994& 1.57375& 0.15782& 0.76938&& 1.79174 &2.53874& 0.02874& 0.59496\\
			&400&1.23585& 1.57783& 0.15538& 0.77398&& 1.82192& 2.52035 &0.02756& 0.60055\\
		
			&&&&&&&&&&\\
			\hline
			&&&&&&&&&&\\
			RTAD&50&1.07211 &0.78706 &0.44628 &0.23619 &&1.56109  &1.14094 &0.99498 &0.10499\\
			&100&0.86686 &0.62946 &0.21951 &0.19058 &&1.07590 &0.75414 &0.11505 &0.0676\\
			&150&0.68424 &0.49725 &0.15114 &0.16721 &&0.70763 &0.46844 &0.03471 &0.04866 \\
			&200&0.60897 &0.42187 &0.12820 &0.14670 &&0.59192 &0.32674 &0.02477 &0.03174 \\
			&300&0.46243 &0.28433 &0.09574 &0.13108 &&0.37528 &0.15427 &0.01360 &0.02242\\
			&400&0.27793 &0.18099 &0.06075 &0.11151 &&0.15612 &0.05862 &0.00596 &0.01449\\
			
			&&&&&&&&&&\\
			\hline 
			\toprule 
		\end{tabular}\label{T1}
	\end{table}	
	
	\begin{table}[h]
		\caption{ Absolute biases and MSEs of the estimators for settings of parameters $\zeta$=(1.5, 2.0, 0.9, 0.8).} 
		\centering 
		\footnotesize
		\begin{tabular}{ccccccccccc} 
			\hline
			\toprule 
			$N=500$&$\downarrow$Sample& \multicolumn{4}{c}{Absolute Bias} &&\multicolumn{2}{c}{\hspace{2.1cm}MSE} &&\\
			\cmidrule{3-6} \cmidrule{8-11}  
			Estimators &(n)& $\hat{\alpha}$& $\hat{\beta}$ &$\hat{\theta}$&$\hat{\lambda}$&&$\hat{\alpha}$& $\hat{\beta}$ &$\hat{\theta}$&$\hat{\lambda}$  \\
			\hline	
			\toprule 
			&&&&&&&&&&\\
			
			ML&50&1.06842& 0.63133 &0.66294& 0.18474&& 2.14481&  1.18595&  1.06159 &0.04188\\
			&100&0.75452 &0.44186& 0.39654& 0.17150 &&1.23565& 0.36786& 0.29582& 0.03780\\
			&150&0.46720 &0.32625& 0.26717& 0.14833&& 0.59076& 0.15978 &0.11982& 0.02910\\
			&200&0.33276& 0.31043& 0.20992& 0.15236&& 0.27355& 0.13543& 0.07756& 0.02980\\
			&300&0.22182 &0.27704& 0.15516& 0.13014&& 0.09038& 0.10674& 0.03846& 0.02228\\
			&400&0.18318& 0.24841& 0.12535& 0.11693&& 0.04863& 0.09014& 0.02281& 0.01929\\

			&&&&&&&&&&\\
			\hline
			&&&&&&&&&&\\
			
			OLS&50&1.53230& 1.14109& 2.13426 &0.23204 &&5.39160&  5.12633 &39.10015& 0.09997\\
			&100&1.27249& 0.83403& 0.82789& 0.21750&& 3.87997& 3.29300 &2.28042& 0.08544\\
			&150&0.94327 &0.58092& 0.56197 &0.18973 &&2.04946 &0.82230& 0.84719 &0.06777\\
			&200&0.70319&0.45971& 0.41839 &0.17813&&1.20894 &0.38355& 0.33944& 0.05582\\
			&300&0.42569& 0.38484& 0.29103& 0.17154&& 0.35235& 0.19071& 0.13794 &0.04148\\
			&400&0.27933 &0.35815& 0.20423& 0.17459&& 0.11932 &0.13946& 0.06450 &0.03318\\

			&&&&&&&&&&\\
			\hline
			&&&&&&&&&&\\
			WLS&50&1.38887 &1.21553& 1.48579 &1.50174 &&4.72591& 19.35346& 19.12901& 0.09997\\
			&100&1.00383 &0.64798& 0.58305& 0.58894&& 2.75266& 0.84834& 0.79954& 0.08544\\
			&150&0.71478& 0.48145& 0.40808 &0.41376 &&1.32532 &0.36813 &0.34054 &0.06777\\
			&200&0.52125& 0.36730& 0.31073 &0.32559 &&0.72584 &0.19564& 0.17507& 0.05582\\
			&300&0.33592& 0.26993 &0.22134& 0.23156&& 0.22198& 0.08993& 0.07760& 0.04148\\
			&400&0.21130& 0.19784 &0.14608& 0.15308 &&0.07287 &0.04215& 0.03434 &0.03318\\

			&&&&&&&&&&\\
			\hline
			&&&&&&&&&&\\
			CvM&50&0.89123& 1.29216 &0.07220& 0.37500&& 0.79474&  1.67301 & 0.01359 &0.14081\\	
			&100&0.89928 &1.28051& 0.04827 &0.37157&& 0.80908& 1.64191& 0.00772 &0.13819\\
			&150&0.90304&1.27433& 0.03504& 0.36992&& 0.81576 &1.62541 &0.00495& 0.13693\\
			&200&0.90571 &1.26928& 0.02626& 0.36863 &&0.82049&1.61195& 0.00288& 0.13594\\
			&300&0.90787 &1.26502 &0.01828& 0.36758 &&0.82432& 1.60061 &0.00118& 0.13513\\
			&400&0.90935& 1.26275& 0.01458 &0.36701&& 0.82693& 1.59453& 0.00025& 0.13470\\

			&&&&&&&&&&\\
			\hline
			&&&&&&&&&&\\
			
			AD&50&0.91197& 0.67772& 0.85022& 0.20946&& 1.56453&  1.26201&  1.90796& 0.08346\\
			&100&0.73953 &0.46551& 0.52012& 0.17600&& 1.05249& 0.41067 &0.64484& 0.05763\\
			&150&0.55563& 0.38209& 0.35936& 0.15859 &&0.62817 &0.26853 &0.26511 &0.04856\\
			&200&0.43489& 0.32129& 0.28029& 0.13541&& 0.41753& 0.17716& 0.15051& 0.03278\\
			&300&0.30437& 0.25573& 0.20243& 0.11295 &&0.17403& 0.09997& 0.06586& 0.02019\\
			&400&0.19880& 0.18847& 0.13630& 0.08958&& 0.06447 &0.05729 &0.02948& 0.01275\\

			&&&&&&&&&&\\
			\hline
			&&&&&&&&&&\\
			
			RTAD&50&0.95347& 0.67013& 1.11546& 0.20439&& 1.63062& 1.24779 &3.58082& 0.07612\\
			&100&0.69700 &0.47975 &0.58381& 0.19714 &&0.98766& 0.45678 &0.99601& 0.06844\\
			&150&0.48059 &0.34499 &0.36841& 0.15429 &&0.47031& 0.19084 &0.24918& 0.04013\\
			&200&0.36237 &0.31445 &0.28426& 0.15545 &&0.27020& 0.15183 &0.14895& 0.03913\\
			&300&0.26788 &0.28002 &0.21205& 0.13592 &&0.11388& 0.11152 &0.07446& 0.02679\\
			&400&0.18756 &0.23731 &0.14741& 0.11814 &&0.05488& 0.07539 &0.03734& 0.01787 \\
			
			&&&&&&&&&&\\
			\hline 
			\toprule 
		\end{tabular}\label{T2}
	\end{table}

	\begin{table}[h]
		\caption{ Absolute biases and MSEs of the estimators for settings of parameters $\zeta$=(2.0,2.5,0.7,0.5).} 
		\centering 
		\footnotesize
		\begin{tabular}{ccccccccccc} 
			\hline
			\toprule 
			$N=500$&$\downarrow$Sample& \multicolumn{4}{c}{Absolute Bias} &&\multicolumn{2}{c}{\hspace{2.1cm}MSE} &&\\
			\cmidrule{3-6} \cmidrule{8-11}  
			Estimators &(n)& $\hat{\alpha}$& $\hat{\beta}$ &$\hat{\theta}$&$\hat{\lambda}$&&$\hat{\alpha}$& $\hat{\beta}$ &$\hat{\theta}$&$\hat{\lambda}$  \\
			\hline	
			\toprule 
			&&&&&&&&&&\\
			
			ML&50&1.15264& 0.99238& 0.49580& 0.24711 &&2.11977 & 2.04722&  0.86321& 0.10954\\
			&100&0.89688 &0.74331 &0.29730& 0.22286 &&1.45355& 1.28990 &0.17810& 0.09157\\
			&150&0.61300 &0.54276 &0.19723 &0.22484 &&0.78022& 0.55272& 0.06615& 0.08657\\
			&200&0.50548& 0.46913& 0.16973& 0.21383&& 0.52221&0.35518& 0.04825& 0.07682\\
			&300&0.38450& 0.40310& 0.13065& 0.20588 &&0.28271&0.22616& 0.02444 &0.06378\\
			&400&0.34620& 0.36613& 0.11819& 0.19609 &&0.18776& 0.17885& 0.01900& 0.05269\\

			&&&&&&&&&&\\
			\hline
			&&&&&&&&&&\\
			
			OLS&50&1.80577& 2.00064& 1.48150 &0.29233 &&6.77693& 24.59193& 15.39365 &0.10110\\
			&100&1.43069 &1.09241& 0.65351 &0.28421&& 4.44634& 5.84342& 1.40197& 0.09694\\
			&150&1.08302& 0.69606& 0.43909& 0.26633 &&2.38514 &1.31463& 0.56619 &0.08589\\
			&200&0.80236 &0.59099 &0.31753& 0.23868&& 1.40432& 0.94454& 0.22543& 0.07156\\
			&300&0.59749 &0.44899 &0.22609 &0.22945 &&0.63159 &0.31865& 0.08320 &0.06829\\
			&400&0.42055 &0.43105 &0.14905 &0.23686&& 0.29443&0.25857 &0.03422 &0.07124\\

			&&&&&&&&&&\\
			\hline
			&&&&&&&&&&\\
			WLS&50&1.50074& 1.67590& 1.04656& 1.09217&& 5.19298&  7.47220 & 7.13104 &0.10110\\
			&100&1.06299& 0.98555 &0.43757& 0.46862&&2.92902 &0.61771&0.51239& 0.09694\\
			&150&0.69489 &0.64075& 0.28576& 0.33009 &&1.08675 &0.25425& 0.18607&0.08589\\
			&200&0.50515 &0.55975& 0.21328& 0.26467 &&0.58053& 0.14555 &0.09357& 0.07156\\
			&300&0.37140 &0.48365 &0.15094& 0.20533&& 0.26376& 0.07181 &0.03644 &0.06829\\
			&400&0.26204& 0.50285 &0.09985& 0.16178&& 0.12165& 0.03797 &0.01552 &0.07124\\

			&&&&&&&&&&\\
			\hline
			&&&&&&&&&&\\
			CvM&50&0.77800& 0.38575 &0.27370 &0.19256 &&0.60736 & 0.18245 & 0.07649& 0.03719\\	
			&100&0.79218 &0.38618& 0.28152 &0.19683&& 0.62793& 0.18334& 0.08001& 0.03876\\
			&150&0.79706 &0.38206 &0.28456 &0.19834&& 0.63543 &0.17834 &0.08134& 0.03934\\
			&200&0.79933& 0.36170 &0.28651 &0.19908 &&0.63898& 0.15211& 0.08233 &0.03963\\
			&300&0.80098 &0.33357 &0.28862 &0.19964&& 0.64158 &0.11975 &0.08343 &0.03986\\
			&400&0.80134& 0.30914& 0.29105 &0.19988 &&0.64215& 0.09644& 0.08477 &0.03995\\

			&&&&&&&&&&\\
			\hline
			&&&&&&&&&&\\
			
			AD&50&1.12503 &0.78020 &0.73127 &0.37296 &&1.91855&  0.96390 & 1.59959 &0.15529\\
			&100&0.99218& 0.78800 &0.44771& 0.39955&& 1.50648& 0.79041 &0.46223 &0.17286\\
			&150&0.81457 &0.78740 &0.33564 &0.41103 &&1.06211 &0.73465 &0.22390 &0.17891\\
			&200&0.65669& 0.83018& 0.26591& 0.42269&& 0.72177& 0.78438& 0.12608& 0.18726\\
			&300&0.52564& 0.88144 &0.20939& 0.43596&& 0.44396& 0.83394 &0.06969 &0.19596\\
			&400&0.37086& 0.93135& 0.14979 &0.45294 &&0.18687 &0.89824& 0.03303 &0.20933\\

			&&&&&&&&&&\\
			\hline
			&&&&&&&&&&\\
			
			RTAD&50&1.01208 &0.69588 &0.83242& 0.26892 &&1.57065 &1.07677 &2.46307 &0.09034\\
			&100& 0.75233 &0.58604 &0.4082& 0.25534&& 0.95486& 0.86838& 0.49842& 0.08364\\
			&150&0.55705 &0.45755 &0.25954 &0.26206&& 0.56630 &0.36883 &0.12693 &0.08569\\
			&200&0.41250 &0.41611& 0.18878& 0.24451&& 0.31563 &0.28316& 0.07033 &0.07874\\
			&300&0.28528 &0.37863 &0.13171 &0.26567&& 0.12880& 0.20729& 0.02845& 0.08851\\
			&400&0.23192 &0.34840 &0.08937 &0.24953&& 0.07843& 0.15925& 0.01268& 0.08165\\

			&&&&&&&&&&\\
			\hline 
			\toprule 
		\end{tabular}\label{T3}
	\end{table}

	\clearpage
	\newpage
	\section{Data Analysis} In this section, we have analyzed two different real data sets to demonstrate the usefulness of the proposed GTLD models which are more desirable and extensively used in real-life modeling. For the versatility of the proposed models, we have taken here some special sub-families of GTLD by considering $g^{\alpha}(x) = x$, $g^{\alpha}(x) = x^\alpha$, $g^{\alpha}(x) = e^{x^\alpha}-1$, and $g^{\alpha}(x)= \log(1+x/\alpha)$ which reduces the GTLD to GTE, GTW, GTWE, and GTL distribution respectively. Considered data sets are also fitted to some well-known distributions namely; Weibull (W), Weibull extension (WE) (see Maswadah (\citeyear{maswadah2022improved})), Kumaraswamy alpha power exponential inverse exponential (KAPIE) (see Thomas et al. (\citeyear{thomas2019kumaraswamy})), Kumaraswamy modified inverse Weibull (KMIW) (see Cordeiro et al. (\citeyear{cordeiro2014kumaraswamy})), Kumaraswamy alpha power lomax (KAPL), and alpha power inverse Weibull (APIW) (see Basheer (\citeyear{basheer2019alpha})) distribution. In the beginning, we performed the exploratory data analysis and computed the MLEs of the considered sub-families of the GTLD with other taken distributions. Then, we have done a comparative study of the considered sub-families of the GTLD with other considered distributions. The comparison is made based on some selection statistics such as the value of -2log-likelihood function of the fitted distribution, Akaike's information criteria (AIC) of the fitted model, and some goodness of fit test statistics; Kolmogorov-Smirnov (KS), Cram{\'e}r von Mises (CvM), and Anderson-Darling (AD) along with their p-values. The smallest value of AIC and goodness of fit statistic (with their highest p-value) provides the better fit of distribution. The functional form of these measures and their performing algorithm are easily available in R-programming library. 
	
\begin{figure}[h]
		\centering
		\subfloat{\includegraphics[scale=0.60, angle=0]{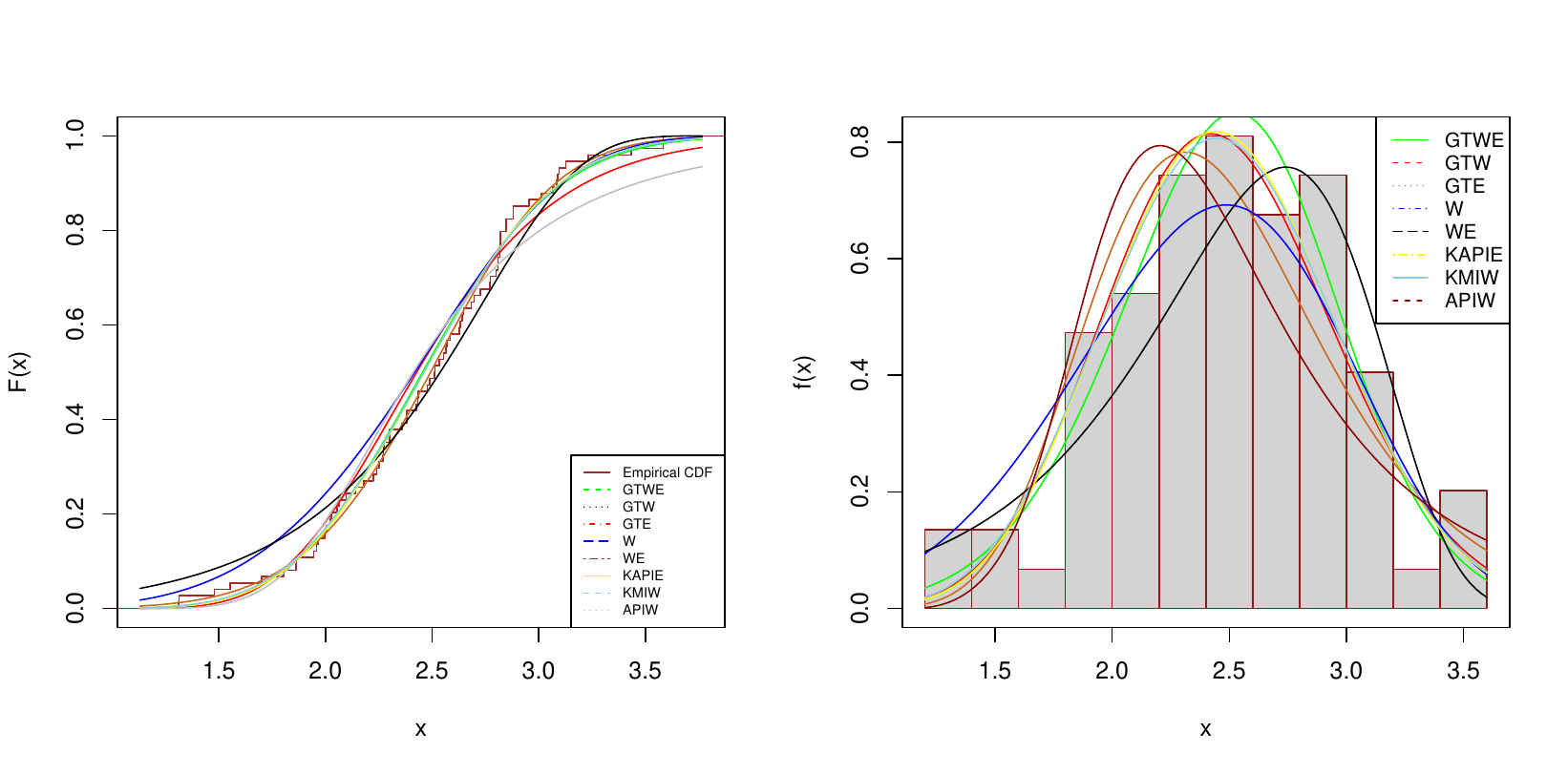}}\\
		\caption{ Fitted CDFs and PDFs for gauge data set.}
		\label{Fig2}
	\end{figure}

	\noindent\textbf{Data Analysis 1:} Here, we have considered the gauge data set (gauge lengths of 20 mm) studied by Kundu and Raqab (\citeyear{kundu2009estimation}) which consists of 74 observations. The data set is as follows: 
	\{1.312, 1.314, 1.479, 1.552, 1.700, 1.803, 1.861, 1.865, 1.944, 1.958, 1.966, 1.997, 2.006, 2.021, 2.027, 2.055, 2.063,
	2.098, 2.140, 2.179, 2.224, 2.240, 2.253, 2.270, 2.272, 2.274, 2.301, 2.301, 2.359, 2.382, 2.382, 2.426, 2.434, 2.435,
	2.478, 2.490, 2.511, 2.514, 2.535, 2.554, 2.566, 2.570, 2.586, 2.629, 2.633, 2.642, 2.648, 2.684, 2.697, 2.726, 2.770,
	2.773, 2.800, 2.809, 2.818, 2.821, 2.848, 2.880, 2.809, 2.818, 2.821, 2.848, 2.880, 2.954, 3.012, 3.067, 3.084, 3.090,
	3.096, 3.128, 3.233, 3.433, 3.585, 3.585\}.
	To better understand the nature of the gauge data set, first, we have obtained some basic statistics and presented in Table \ref{T6}.
	\begin{table}[ht]
		\caption{ The descriptive statistic for gauge datasets.} 
		\centering 
		\footnotesize
		\begin{tabular}{cccccccc} 
			\hline
			Minimum & 1st Quartile&  Median&    Mean& 3rd Quartile&  Maximum&Skewness&Kurtosis   \\
			1.312  & 2.150   &2.513   &2.477  & 2.816 &  3.585 &-0.157396 &0.03344725\\
			\hline	
		\end{tabular}\label{T6}
	\end{table}
	The basic information provided in Table \ref{T6} indicates that the considered data set is negatively skewed, has low kurtosis, and has similar mean and median values. For modeling this negatively skewed and low kurtosis data a distribution having good features is needed. Therefore, we have fitted the data set with respect to GTE, GTW, GTWE, W, WE, KAPIE, KMIW, and APIW distribution and calculated MLEs with their standard errors that are reported in Table \ref{T7}. Furthermore, the fitted empirical and theoretical CDFs and PDFs plots using MLEs are shown in Figure \ref{Fig2}. These fitted figures reveals a good fit of gauge data set with GTWE distribution. As well as, we have calculated the values of $-2 \log L$ and $AIC = 2k-2\log L$, where $k$ is the number of parameters and $L$ denotes the maximized value of the likelihood function. In addition to find the applicability of GTWE distribution, three other selection statistics, namely, KS, CvM, and AD along with their corresponding p-values are also computed and reported in Table \ref{T8}. The smallest value of AIC and goodness of fit statistics (with their high p-value) supports the best fit of GTWE distribution among the considered distributions.

	\begin{table}[h]
		\caption{ Parameter estimates with standard error* values.} 
		\centering 
		\footnotesize
		\begin{tabular}{cccccc} 
			\toprule
			Distribution &$\hat{\alpha}$ &$\hat{\beta}$&$\hat{\theta}$& $\hat{\lambda}$&$\hat{\phi}$ \\
			\toprule
			GTWE& 1.056 (0.174*)&0.108(0.083*) &3.641(2.035*) & 0.669(0.576*)&-\\
			GTW& 3.099(1.348*)&0.100(0.132*) & 3.370(3.396*) &  0.100(0.808*)&-\\
			GTE& -&1.484(0.072*) & 44.259(8.535*) &0.990(0.345*)&-\\
			W&4.799(0.652*) &0.010(0.007*)&-&-&-\\
			WE&1.380(0.051*)&0.019(0.005*) & -&-&-\\
			KAPIE&0.010(0.206*)&118.98(221.24*) &33.914(2.831*) & 0.421(0.019*)&-\\
			KMIW&0.373(0.202*)& 340.61(431.48*) &0.923(0.721*) & 0.011(452.70*)&38.18(452.28*)\\
			APIW&108.95(137.07*)&5.383(0.433*) &18.060 (7.210*)&-&-\\
			\hline
		\end{tabular}\label{T7}
	\end{table}

\begin{figure}[h]
		\centering
		\subfloat{\includegraphics[scale=0.60, angle=0]{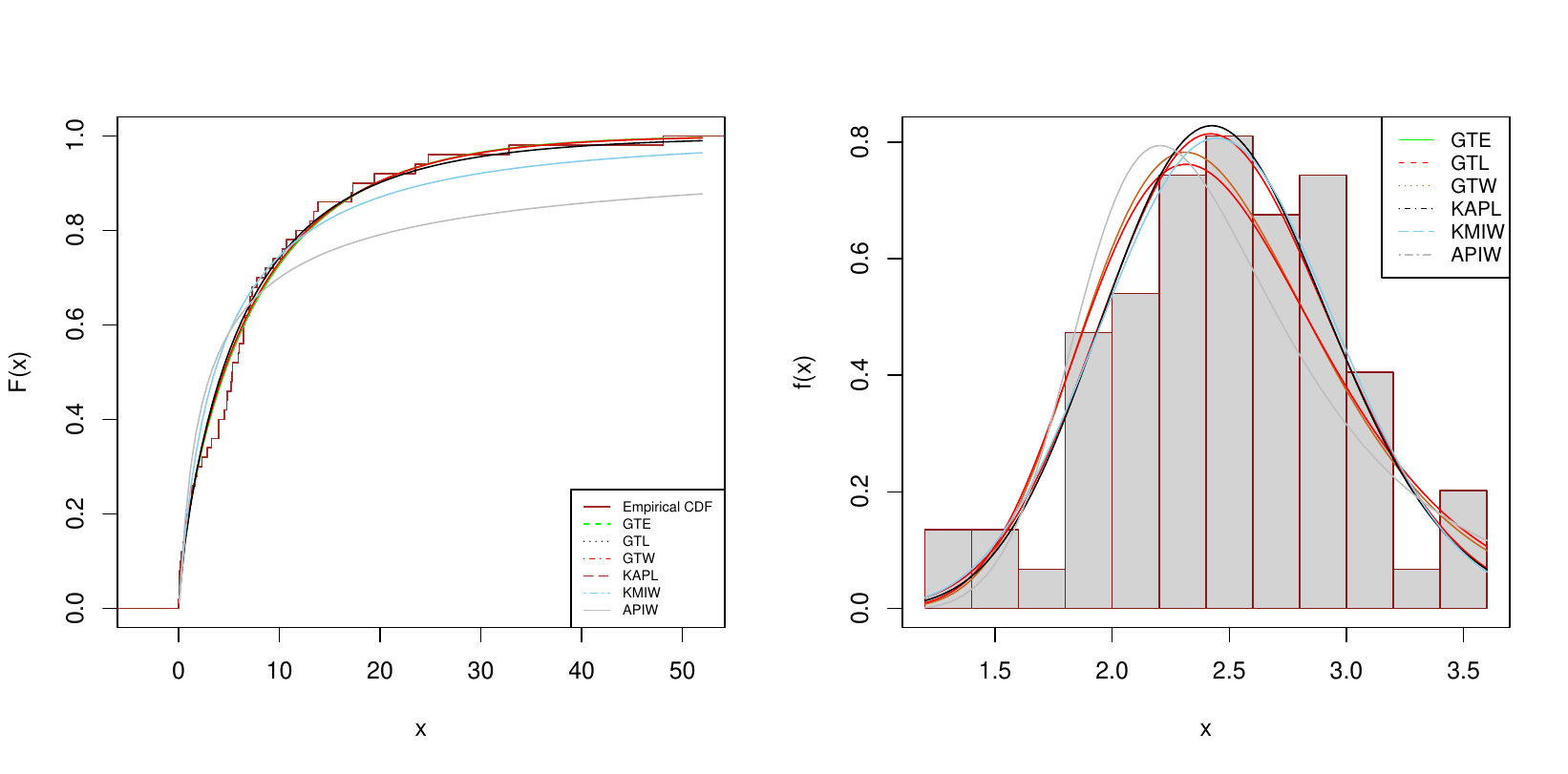}}\\
		\caption{ Fitted CDFs and PDFs for failure time data set.}
		\label{Fig3}
	\end{figure}

	\begin{table}[ht]
		\caption{ Model selection statistics for gauge data set.} 
		\centering 
		\footnotesize
		\begin{tabular}{ccccccccccc} 
			
			\hline
			\toprule 
			$n=74$&& \multicolumn{3}{c}{\hspace{1.6cm}KS} &&\multicolumn{2}{c}{CvM}&& \multicolumn{2}{c}{AD}  \\
			\cmidrule{4-5} \cmidrule{7-8} \cmidrule{10-11} 
			Model &$-2\log L$ & AIC&  Statistic &P-value&& Statistic &P-value&& Statistic &P-value  \\
			\hline	
			&&&&&&&&&&\\
			GTWE&102.2021& 110.2021& 0.05132& 0.98989&& 0.02578 &0.98838 &&0.19068& 0.99277\\
			GTW&103.9872& 111.9872 &0.05841& 0.96236&& 0.06223 &0.80146&& 0.38986 &0.85832\\
			GTE&110.6403& 116.6403& 0.07970& 0.73504 &&0.13295& 0.44685 &&0.89284& 0.41831\\
			W&106.9877& 110.9877& 0.10757& 0.35869&& 0.23601& 0.20746 &&1.34764& 0.21740\\
			WE&113.3503& 117.3503& 0.09847& 0.46975 &&0.12922 &0.46051&& 1.06990& 0.32236\\
			KAPIE& 103.6176 &111.6176& 0.05714 &0.96909&& 0.05077& 0.87302&& 0.34771& 0.89819\\
			KMIW&103.3950 &113.3950& 0.05543 &0.97687 &&0.05244& 0.86278&& 0.34353& 0.90194\\
			APIW&123.1184 &129.1184& 0.11154 &0.31598&& 0.22775& 0.21974 &&1.71132 &0.13331\\	 	
			\hline 
		\end{tabular}\label{T8}
	\end{table}	
	\clearpage
	\noindent \textbf{Data Analysis 2:} Another data set that we have considered here is a failure times (in weeks) data set of 50 components (see Tanis and Saracoglu (\citeyear{tanics2022record})). The observations of considered data are as follows: \{0.013, 0.065, 0.111, 0.111, 0.163, 0.309, 0.426, 0.535, 0.684, 0.747, 0.997, 1.284, 1.304, 1.647, 1.829, 2.336, 2.838, 3.269, 3.977, 3.981, 4.520, 4.789, 4.849, 5.202, 5.291, 5.349, 5.911, 6.018, 6.427, 6.456, 6.572, 7.023, 7.087, 7.291, 7.787, 8.596, 9.388, 10.261, 10.713, 11.658, 13.006, 13.388, 13.842, 17.152, 17.283, 19.418, 23.471, 24.777, 32.795, 48.105\}. 
	We have performed a similar analysis for this data set same as in gauge data with respect to GTE, GTL, GTW, KAPL, KMIW, and APIW distribution. All the results are reported in Table \ref{T9}-\ref{T11}. We found that the GTE distribution fits this data better than the GTL, GTW, KAPL, KMIW, and APIW distribution in terms of lowest values of AIC and goodness of fit statistics (with their high p-values) as shown in Figure \ref{Fig3} and Table \ref{T11}, respectively.  
	\begin{table}[h]
		\caption{ The descriptive statistic for failure time  datasets.} 
		\centering 
		\footnotesize
		\begin{tabular}{cccccccc} 
			\hline
			Minimum & 1st Quartile&  Median&    Mean& 3rd Quartile&  Maximum&Skewness&Kurtosis   \\
			0.013  & 1.390  & 5.320  & 7.821 & 10.043&  48.105& 2.377991 &7.228855\\
			\hline	
		\end{tabular}\label{T9}
	\end{table}	
	\begin{table}[h]
		\caption{Parameter estimates with standard error* values.} 
		\centering
		\footnotesize
		\begin{tabular}{cccccc} 
			\toprule
			Distribution &$\hat{\alpha}$ &$\hat{\beta}$&$\hat{\theta}$& $\hat{\lambda}$&$\hat{\phi}$ \\
			\toprule
			GTE& -& 0.099 (0.038*) & 0.688 (0.230*) &0.01 (0.904*)&-\\
			GTL& 41237.36 (142.007*)&4095.918 (1547.409*) &0.690 (0.226*) & 0.01 (0.883*)&-\\
			GTW& 0.991 (0.247*)&0.100 (0.138*)& 0.721 (0.267*) &  0.1 (0.633*)&-\\
			KAPL&0.01 (0.101*)& 5.271 (15.129*)& 0.01 (0.011*)& 0.801 (0.13*)& 0.392 (1.129*)\\
			KMIW&0.018 (0.008*) &53.159 (38.046*)& 0.164 (0.029*)& 0.01 (1.37*)& 298.55 (126.59*)\\
			APIW&102.38 (161.044*) &0.622 (0.059*)& 0.332 (0.121*)&-&-\\
			\hline
		\end{tabular}\label{T10}
	\end{table}	
	\begin{table}[ht]
		\caption{ Model evaluation statistics for failure time data set.} 
		\centering 
		\footnotesize
		\begin{tabular}{ccccccccccc}
			\hline
			\toprule 
			$n=50$&& \multicolumn{3}{c}{\hspace{1.6cm}KS} &&\multicolumn{2}{c}{CvM}&& \multicolumn{2}{c}{AD}  \\
			\cmidrule{4-5} \cmidrule{7-8} \cmidrule{10-11} 
			Model &$-2\log L$ & AIC&  Statistic &P-value&& Statistic &P-value&& Statistic &P-value  \\
			\hline	
			&&&&&&&&&&\\
			GTE&300.6038& 306.6038& 0.10368& 0.65559&& 0.06591& 0.77916 &&0.32867 &0.91478\\
			GTL&300.6047& 308.6047& 0.10411 &0.65048 &&0.06628 &0.77683&& 0.33018 &0.91348\\
			GTW&300.7535 &308.7535& 0.10919 &0.59011&& 0.07345& 0.73279 &&0.36132 &0.88552\\
			KAPL& 301.7533 &311.7533& 0.11876& 0.48104& &0.09429& 0.61551 &&0.48252& 0.76390\\
			KMIW&309.7273& 319.7273& 0.16629& 0.12588&& 0.22987& 0.21659 &&1.24830 &0.24970\\
			APIW&326.8694 &332.8694 &0.18033& 0.07740 &&0.40208 &0.07097 &&2.40525& 0.05582\\	 	
			\hline 
		\end{tabular}\label{T11}
	\end{table}	
	
	
	\section{Conclusion}
	
	In this article, we have introduced a generalized transmuted class of lifetime distribution and called it GTLD. The proposed GTLD includes a generalized transmuted exponential (GTE), generalized transmuted Rayleigh (GTR), generalized transmuted Weibull (GTW), generalized transmuted modified Weibull (GTMW), generalized transmuted Weibull extension (GTWE), generalized transmuted Burr-type-XII (GTB-XII), generalized transmuted Lomax (GTL), and generalized transmuted pareto type-I (GTP-I) distribution as a sub-family. Several mathematical quantities namely; density function, distribution function, quantile function, various moments, moment generating function, stress-strength reliability function, order statistics, R{\'e}nyi and q-entropy,  residual and reversed residual life function, and cumulative information generating function (CIGF) are also discussed. Apart from that, for estimation of unknown parameters, maximum likelihood (ML), ordinary least square (OLS), weighted least square (WLS), Cram{\'e}r-von Mises (CvM), Anderson Darling (AD), and Right-tail Anderson Darling (RTAD) methods have been considered. Using Monte Carlo algorithm, a well-organized simulation study is performed to observe the behavior of estimators under the absolute bias and the mean squared error criteria. Finally, the analysis of two real data sets have also been presented which suggest that the special sub-family of the GTLD  can provide better fit as compared to the other models taken from well known families of the distributions.  \\
	
	\noindent\textbf{Author's Contributions:} All authors contributed equally in this paper.\\
	\noindent\textbf{Funding:} No funding.\\
	\noindent\textbf{Availability of Data and Materials:} Not applicable.\\
	
	\noindent\textbf{Declarations:}\\
	
	\noindent\textbf{Ethical Approval:} Not required.\\
	\noindent\textbf{Competing Interests:} Not applicable.\\
	\noindent\textbf{Disclosure statement:}
	No potential conflict of interest was reported by the author(s).
	\bibliographystyle{apalike}
	\bibliography{Alok.bib}

\begin{thebibliography}{}

\bibitem[Alizadeh et~al., 2017a]{alizadeh2017new}
Alizadeh, M., Fazel~Bagheri, S., Alizadeh, M., and Nadarajah, S. (2017a).
\newblock A new four-parameter lifetime distribution.
\newblock {\em Journal of Applied Statistics}, 44(5):767--797.

\bibitem[Alizadeh et~al., 2017b]{alizadeh2017generalized}
Alizadeh, M., Merovci, F., and Hamedani, G. (2017b).
\newblock Generalized transmuted family of distributions: properties and
  applications.
\newblock {\em Hacettepe Journal of Mathematics and Statistics},
  46(4):645--667.

\bibitem[Alzaatreh et~al., 2013]{alzaatreh2013new}
Alzaatreh, A., Lee, C., and Famoye, F. (2013).
\newblock A new method for generating families of continuous distributions.
\newblock {\em Metron}, 71(1):63--79.

\bibitem[Arshad et~al., 2023]{arshad2023bayesian}
Arshad, M., J.~Azhad, Q., Gupta, N., and Pathak, A.~K. (2023).
\newblock Bayesian inference of unit gompertz distribution based on dual
  generalized order statistics.
\newblock {\em Communications in Statistics-Simulation and Computation},
  52(8):3657--3675.

\bibitem[Arshad et~al., 2022]{arshad2022record}
Arshad, M., Khetan, M., Kumar, V., and Pathak, A.~K. (2022).
\newblock Record-based transmuted generalized linear exponential distribution
  with increasing, decreasing and bathtub shaped failure rates.
\newblock {\em Communications in Statistics-Simulation and Computation}, pages
  1--25.

\bibitem[Aryal and Tsokos, 2009]{aryal2009transmuted}
Aryal, G.~R. and Tsokos, C.~P. (2009).
\newblock On the transmuted extreme value distribution with application.
\newblock {\em Nonlinear Analysis: Theory, Methods \& Applications},
  71(12):e1401--e1407.

\bibitem[Aryal and Tsokos, 2011]{aryal2011transmuted}
Aryal, G.~R. and Tsokos, C.~P. (2011).
\newblock Transmuted weibull distribution: A generalization of theweibull
  probability distribution.
\newblock {\em European Journal of pure and applied mathematics}, 4(2):89--102.

\bibitem[Azhad et~al., 2021]{azhad2021estimation}
Azhad, Q.~J., Arshad, M., and Misra, A.~K. (2021).
\newblock Estimation of common location parameter of several heterogeneous
  exponential populations based on generalized order statistics.
\newblock {\em Journal of Applied Statistics}, 48(10):1798--1815.

\bibitem[Basheer, 2019]{basheer2019alpha}
Basheer, A.~M. (2019).
\newblock Alpha power inverse weibull distribution with reliability
  application.
\newblock {\em Journal of Taibah University for Science}, 13(1):423--432.

\bibitem[Bhatti et~al., 2018]{bhatti2018transmuted}
Bhatti, F.~A., Hamedani, G., Korkmaz, M.~{\c{C}}., and Ahmad, M. (2018).
\newblock The transmuted geometric-quadratic hazard rate distribution:
  development, properties, characterizations and applications.
\newblock {\em Journal of Statistical Distributions and Applications}, 5:1--23.

\bibitem[Broyden, 1970]{broyden1970convergence}
Broyden, C.~G. (1970).
\newblock The convergence of a class of double-rank minimization algorithms: 2.
  the new algorithm.
\newblock {\em IMA journal of applied mathematics}, 6(3):222--231.

\bibitem[Capaldo et~al., 2023]{capaldo2023cumulative}
Capaldo, M., Di~Crescenzo, A., and Meoli, A. (2023).
\newblock Cumulative information generating function and generalized gini
  functions.
\newblock {\em Metrika}, pages 1--29.

\bibitem[Cordeiro and de~Castro, 2011]{cordeiro2011new}
Cordeiro, G.~M. and de~Castro, M. (2011).
\newblock A new family of generalized distributions.
\newblock {\em Journal of statistical computation and simulation},
  81(7):883--898.

\bibitem[Cordeiro et~al., 2013]{cordeiro2013exponentiated}
Cordeiro, G.~M., Ortega, E.~M., and da~Cunha, D.~C. (2013).
\newblock The exponentiated generalized class of distributions.
\newblock {\em Journal of data science}, 11(1):1--27.

\bibitem[Cordeiro et~al., 2014]{cordeiro2014kumaraswamy}
Cordeiro, G.~M., Ortega, E.~M., and Silva, G.~O. (2014).
\newblock The kumaraswamy modified weibull distribution: theory and
  applications.
\newblock {\em Journal of Statistical Computation and Simulation},
  84(7):1387--1411.

\bibitem[Dey et~al., 2018a]{dey2018kumaraswamy}
Dey, S., Mazucheli, J., and Nadarajah, S. (2018a).
\newblock Kumaraswamy distribution: different methods of estimation.
\newblock {\em Computational and Applied Mathematics}, 37:2094--2111.

\bibitem[Dey et~al., 2018b]{dey2018statistical}
Dey, S., Moala, F.~A., and Kumar, D. (2018b).
\newblock Statistical properties and different methods of estimation of
  gompertz distribution with application.
\newblock {\em Journal of Statistics and Management Systems}, 21(5):839--876.

\bibitem[Elbatal, 2013]{elbatal2013transmuted}
Elbatal, I. (2013).
\newblock Transmuted generalized inverted exponential distribution.
\newblock {\em Economic Quality Control}, 28(2):125--133.

\bibitem[Fletcher, 1970]{fletcher1970new}
Fletcher, R. (1970).
\newblock A new approach to variable metric algorithms.
\newblock {\em The computer journal}, 13(3):317--322.

\bibitem[Goldfarb, 1970]{goldfarb1970family}
Goldfarb, D. (1970).
\newblock A family of variable-metric methods derived by variational means.
\newblock {\em Mathematics of computation}, 24(109):23--26.

\bibitem[Gradshteyn and Ryzhik, 2014]{gradshteyn2014table}
Gradshteyn, I.~S. and Ryzhik, I.~M. (2014).
\newblock {\em Table of integrals, series, and products}.
\newblock Academic press.

\bibitem[Granzotto et~al., 2017]{granzotto2017cubic}
Granzotto, D., Louzada, F., and Balakrishnan, N. (2017).
\newblock Cubic rank transmuted distributions: inferential issues and
  applications.
\newblock {\em Journal of statistical Computation and Simulation},
  87(14):2760--2778.

\bibitem[Granzotto and Louzada, 2015]{granzotto2015transmuted}
Granzotto, D. C.~T. and Louzada, F. (2015).
\newblock The transmuted log-logistic distribution: modeling, inference, and an
  application to a polled tabapua race time up to first calving data.
\newblock {\em Communications in Statistics-Theory and Methods},
  44(16):3387--3402.

\bibitem[Gupta and Kundu, 1999]{gupta1999theory}
Gupta, R.~D. and Kundu, D. (1999).
\newblock Theory \& methods: Generalized exponential distributions.
\newblock {\em Australian \& New Zealand Journal of Statistics},
  41(2):173--188.

\bibitem[Gupter et~al., 1998]{gupter1998modeling}
Gupter, R., Gupta, P., and Gupta, R. (1998).
\newblock Modeling failure time data by lehmann alternatives.
\newblock {\em Communications in Statistics Theory and Methods}, 27:887--904.

\bibitem[Kemaloglu and Yilmaz, 2017]{kemaloglu2017transmuted}
Kemaloglu, S.~A. and Yilmaz, M. (2017).
\newblock Transmuted two-parameter lindley distribution.
\newblock {\em Communications in Statistics-Theory and Methods},
  46(23):11866--11879.

\bibitem[Khan and King, 2013]{khan2013transmuted}
Khan, M.~S. and King, R. (2013).
\newblock Transmuted modified weibull distribution: A generalization of the
  modified weibull probability distribution.
\newblock {\em European Journal of pure and applied mathematics}, 6(1):66--88.

\bibitem[Khan and King, 2014]{khan2014new}
Khan, M.~S. and King, R. (2014).
\newblock A new class of transmuted inverse weibull distribution for
  reliability analysis.
\newblock {\em American Journal of Mathematical and Management Sciences},
  33(4):261--286.

\bibitem[Kundu and Raqab, 2009]{kundu2009estimation}
Kundu, D. and Raqab, M.~Z. (2009).
\newblock Estimation of r= p (y< x) for three-parameter weibull distribution.
\newblock {\em Statistics \& Probability Letters}, 79(17):1839--1846.

\bibitem[Maswadah, 2022]{maswadah2022improved}
Maswadah, M. (2022).
\newblock Improved maximum likelihood estimation of the shape-scale family
  based on the generalized progressive hybrid censoring scheme.
\newblock {\em Journal of Applied Statistics}, 49(11):2825--2844.

\bibitem[Merovci et~al., 2017]{merovci2017exponentiated}
Merovci, F., Alizadeh, M., Yousof, H.~M., and Hamedani, G. (2017).
\newblock The exponentiated transmuted-g family of distributions: theory and
  applications.
\newblock {\em Communications in Statistics-Theory and Methods},
  46(21):10800--10822.

\bibitem[Merovci et~al., 2013]{merovci2013transmuted}
Merovci, F., Elbatal, I., and Ahmed, A. (2013).
\newblock Transmuted generalized inverse weibull distribution.
\newblock {\em arXiv preprint arXiv:1309.3268}.

\bibitem[Nadarajah and Kotz, 2006]{nadarajah2006exponentiated}
Nadarajah, S. and Kotz, S. (2006).
\newblock The exponentiated type distributions.
\newblock {\em Acta Applicandae Mathematica}, 92:97--111.

\bibitem[Nadarajah et~al., 2014]{nadarajah2014new}
Nadarajah, S., Shahsanaei, F., and Rezaei, S. (2014).
\newblock A new four-parameter lifetime distribution.
\newblock {\em Journal of Statistical Computation and Simulation},
  84(2):248--263.

\bibitem[Nofal et~al., 2017]{nofal2017generalized}
Nofal, Z.~M., Afify, A.~Z., Yousof, H.~M., and Cordeiro, G.~M. (2017).
\newblock The generalized transmuted-g family of distributions.
\newblock {\em Communications in Statistics-Theory and Methods},
  46(8):4119--4136.

\bibitem[Saboor et~al., 2016]{saboor2016transmuted}
Saboor, A., Elbatal, I., and Cordeiro, G.~M. (2016).
\newblock The transmuted exponentiated weibull geometric distribution: Theory
  and applications.
\newblock {\em Hacettepe Journal of Mathematics and Statistics},
  45(3):973--987.

\bibitem[SARA{\c{C}}O{\u{G}}LU and TANI{\c{S}}, 2021]{saraccouglu2021new}
SARA{\c{C}}O{\u{G}}LU, B. and TANI{\c{S}}, C. (2021).
\newblock A new lifetime distribution: transmuted exponential power
  distribution.
\newblock {\em Communications Faculty of Sciences University of Ankara Series
  A1 Mathematics and Statistics}, 70(1):1--14.

\bibitem[Shanno, 1970]{shanno1970conditioning}
Shanno, D.~F. (1970).
\newblock Conditioning of quasi-newton methods for function minimization.
\newblock {\em Mathematics of computation}, 24(111):647--656.

\bibitem[Shaw and Buckley, 2009]{shaw2009alchemy}
Shaw, W.~T. and Buckley, I.~R. (2009).
\newblock The alchemy of probability distributions: beyond gram-charlier
  expansions, and a skew-kurtotic-normal distribution from a rank transmutation
  map.
\newblock {\em arXiv preprint arXiv:0901.0434}.

\bibitem[Sousa-Ferreira et~al., 2023]{sousa2023extended}
Sousa-Ferreira, I., Abreu, A.~M., and Rocha, C. (2023).
\newblock The extended chen-poisson lifetime distribution.
\newblock {\em REVSTAT-Statistical Journal}, 21(2):173--196.

\bibitem[Tan{\i}{\c{s}} and Sara{\c{c}}o{\u{g}}lu, 2022]{tanics2022record}
Tan{\i}{\c{s}}, C. and Sara{\c{c}}o{\u{g}}lu, B. (2022).
\newblock On the record-based transmuted model of balakrishnan and he based on
  weibull distribution.
\newblock {\em Communications in Statistics-Simulation and Computation},
  51(8):4204--4224.

\bibitem[Tani{\c{s}} and Sara{\c{c}}o{\u{g}}lu, 2023]{tanics2023cubic}
Tani{\c{s}}, C. and Sara{\c{c}}o{\u{g}}lu, B. (2023).
\newblock Cubic rank transmuted generalized gompertz distribution: properties
  and applications.
\newblock {\em Journal of Applied Statistics}, 50(1):195--213.

\bibitem[TANI{\c{S}} et~al., 2020]{tanics2020transmuted}
TANI{\c{S}}, C., SARA{\c{C}}O{\u{G}}LU, B., Co{\c{s}}kun, K., and PEKG{\"O}R,
  A. (2020).
\newblock Transmuted complementary exponential power distribution: properties
  and applications.
\newblock {\em Cumhuriyet Science Journal}, 41(2):419--432.

\bibitem[THOMAS et~al., 2019]{thomas2019kumaraswamy}
THOMAS, J., Zelibe, S.~C., and Eyefia, E. (2019).
\newblock Kumaraswamy alpha power inverted exponential distribution: properties
  and applications.
\newblock {\em Istatistik Journal of the Turkish Statistical Association},
  12(1):35--48.

\bibitem[Tian et~al., 2014]{tian2014transmuted}
Tian, Y., Tian, M., and Zhu, Q. (2014).
\newblock Transmuted linear exponential distribution: A new generalization of
  the linear exponential distribution.
\newblock {\em Communications in Statistics-Simulation and Computation},
  43(10):2661--2677.

\end{thebibliography}
	
\end{document}